\documentclass[12pt]{article}
\usepackage[utf8]{inputenc}
\usepackage{textcomp}
\usepackage{comment}

\usepackage[section]{placeins}
\usepackage{graphicx}
\usepackage{capt-of} 
\usepackage{amsmath}
\usepackage{amsfonts}
\usepackage[version=4]{mhchem}
\usepackage{siunitx}
\usepackage{longtable,tabularx}
\usepackage{caption}
\usepackage{subcaption}
\usepackage{float}

\DeclareMathOperator{\arccot}{arccot}

\newcommand{\vv}{{\mathbf v}}               

\newcommand{\nn}{{\mathbf n}}               

\newcommand{\ttt}{\mbox{\boldmath$\theta$}} 

\newcommand{\bbb}{{\mathbf b}}               

\usepackage{color}       
\usepackage{ifthen}      
\usepackage{ifthen}      

\newcommand{\mydrafttext}{}
\newcommand{\drafttext}[1]{\renewcommand{\mydrafttext}{#1}}
\newboolean{draft}  
\setboolean{draft}{true}
\ifthenelse{\boolean{draft}}
{
    \newcounter{comments}
    \drafttext{{\color{red}This document is a draft.}}
    \newcommand{\shane}[1]{\addtocounter{comments}{1}{\color{red}\bf [Shane comment \thecomments: #1]}}
    \newcommand{\mohamed}[1]{\addtocounter{comments}{1}{\color{blue}\bf [Mohamed comment \thecomments: #1]}}
    }
{
\newcommand{\shane}[1]{}
\newcommand{\mohamed}[1]{}
}

\title{Global Phase-Space Geometry of Three-Dimensional Gliding:\\ Terminal Velocity Manifolds,\\ Separatrices, and Stability Structure}

\author{Mohamed Zakaria\footnote{PhD Student, Aerospace and Ocean Engineering, Virginia Tech} 
and 
Shane D.\ Ross\footnote{Professor, Aerospace and Ocean Engineering, Virginia Tech}}

\begin{document}

\maketitle

\begin{abstract}

We develop a three-dimensional dynamical-systems framework for passive gliding and identify the global phase-space structures that organize its motion. 
Extending previous two-dimensional models of non-equilibrium gliding, we show that the 3D velocity dynamics possess an attracting, normally hyperbolic invariant surface, the terminal velocity manifold (TVM), onto which all trajectories rapidly collapse before evolving slowly toward a glide equilibrium.
There is also a separatrix surface associated with an invariant manifold of an unstable equilibrium within the TVM, which partitions initial conditions into qualitatively distinct descent behaviors: efficient shallow glides versus steep, drag-dominated descent.
Using lift-drag data from three representative airfoils---a snake-inspired bluff body, the Zimmerman planform characteristic of {\it Draco} lizards, and the classical NACA 0012---we compute the full equilibrium surfaces, analyze their pitch-roll bifurcations, and reconstruct the TVM and separatrix geometry in three dimensions. The results reveal that (i) equilibrium stability changes with both pitch and roll, rather than pitch alone; (ii) separatrix geometry determines the dynamic accessibility of shallow glides; and (iii) bio-inspired airfoils possess compact separatrix regions that make efficient gliding robust across a wide range of initial jump conditions.
This work unifies biological and engineered gliders within a single global geometric framework and establishes separatrix geometry on the TVM as a principled diagnostic for glide robustness.

\end{abstract}

\section{Introduction}

Gliding and directed aerial descent occur across a remarkable diversity of animal taxa, from flying lizards and snakes to ants, frogs, and mammals \cite{socha2015animals,khandelwal2023convergence}.
Despite large differences in morphology, a common theme is the use of body posture and aerodynamic shape to generate stabilizing forces without continuous flapping. 
These passive mechanisms inspire small-scale aerial robots, where exploiting natural stability and low-energy descent can be advantageous \cite{hernandez2023design,shin2018bio}.

To understand such behaviors, dynamical models provide an essential perspective. 
Classical analyses of bodies in steady fluids, such as phugoid dynamics and vortex-induced motion \cite{andronov2013theory,aref1993chaotic}, capture fundamental energy exchanges but typically constrain the motion to two dimensions. 
More recent work by Yeaton et al.\ \cite{yeaton2017global} and Nave \& Ross \cite{nave2019global} introduced a simplified two-dimensional model for non-equilibrium gliding based on fixed body orientation. 
A key discovery in these studies was the presence of a single attracting curve in velocity space onto which all trajectories collapse before converging to equilibrium glide. 
This curve acts as a one-dimensional terminal velocity manifold that organizes the flow.

\begin{figure}[!h]
\centering
\includegraphics[width=\textwidth]{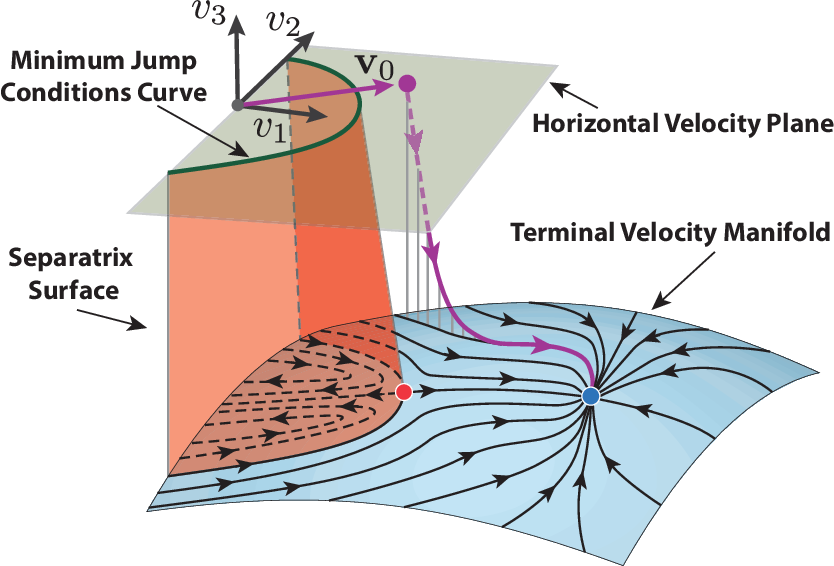}
\caption{{\bf Schematic of the global geometry of 3D gliding.}
Trajectories collapse rapidly onto the terminal velocity manifold (blue), where a separatrix surface (red) partitions initial conditions leading to shallow, lift-dominated glides (blue dot equilibrium) from those that fall into steep, drag-dominated descent. The intersection of the separatrix with the horizontal-velocity plane defines the minimum jump conditions required to reach an efficient glide.}
\label{TVM_separatrix_schematic}
\end{figure}

\paragraph{From 2D to 3D: Global Geometry Becomes Richer.}
In this work, we extend that framework to three dimensions and show that the velocity dynamics admit a two-dimensional attracting invariant manifold, which we call the terminal velocity manifold (TVM).
As illustrated schematically in Figure \ref{TVM_separatrix_schematic}, trajectories collapse rapidly onto the TVM (blue surface) and then evolve slowly along it, reflecting a strong separation of timescales characteristic of normally hyperbolic invariant manifolds.
Crucially, the TVM contains not only the stable equilibrium glide state but also the unstable saddle-like equilibrium whose two-dimensional stable manifold forms a separatrix surface. 
This separatrix acts as a codimension-one barrier that sharply divides initial conditions leading to shallow, lift-dominated glides from those that enter steep descent (Figure \ref{TVM_separatrix_schematic}, red surface).

This geometry, TVM plus separatrix, forms the global ``skeleton'' of gliding dynamics. It determines which equilibrium states are dynamically accessible and how sensitive gliding is to initial velocity orientation.

\paragraph{Airfoil-Dependent Phase-Space Geometry.}
To investigate how biological and engineered shapes exploit this global geometry, we examine three representative cross-sectional airfoils:
\begin{itemize}
	\item a snake-inspired bluff body based on {\it Chrysopelea paradisi} \cite{holden2014aerodynamics},
	\item the Zimmerman planform characteristic of Draco lizards \cite{hassanalian2017design,khandelwal2022combined},
    \item and the classical NACA 0012 airfoil \cite{abbott1949theory}.
\end{itemize}
These profiles allow us to compare (i) glide equilibria and stability; (ii)  bifurcations with respect pitch, yaw, and roll; (iii) the geometry of the two-dimensional TVM; and (iv) the co-dimension-one separatrix geometry that partitions shallow from steep glides.
The differences between these airfoils reveal commonalities and differences in glide behavior across biological and engineered shapes.

\paragraph{Contributions of this Work.}
The key contributions of this study are:
\begin{enumerate}
    \item {\it 3D generalization of the terminal velocity manifold.}
We demonstrate that a 2D attracting invariant surface persists across all airfoils tested and over large ranges of body orientation.
	\item {\it Discovery of a 3D separatrix surface on the TVM.}
This separatrix forms a global transport barrier dividing shallow, efficient glides from steep, drag-dominated descent.
	\item {\it Equilibrium geometry in full pitch-yaw-roll space.}
Stability transitions occur not only in pitch but also under roll and yaw variation, producing rich multistability behavior.
	\item {\it Comparative glide robustness.}
The two bio-inspired airfoils considered have compact separatrix regions, allowing easier access to efficient glides, whereas the engineered NACA 0012 exhibits large separatrix regions and inaccessible equilibria.
\end{enumerate}

We begin in Section 2 by introducing the generalized three-dimensional model, including its derivation from first principles. 
Section 3 examines the equilibrium points of the system and their bifurcations as pitch, roll, and yaw are varied. 
Section 4 develops numerical methods for computing the TVM and separatrix.
Section 5 applies this framework to explore how small roll perturbations reshape the global phase-space structure of the Zimmerman airfoil. Together, these results reveal how global geometric structures govern glide performance in both biological and engineered systems.

\section{Three-Dimensional Glider Model}

We now extend the two-dimensional glider model of \cite{yeaton2017global,nave2019global} into three dimensions by incorporating the full spatial orientation of the body. 
The glider is treated as a rigid body translating under the influence of lift, drag, and gravity, with its aerodynamic forces determined by the instantaneous direction of motion and a prescribed body orientation. 
In this formulation, pitch, roll, and yaw angles are specified parameters rather than dynamical variables; this approach allows us to examine how changes in body attitude alter the translational dynamics without introducing the additional complexity of rotational equations of motion.

\begin{figure}[!h]
\centering
\begin{tabular}{ c c }
\includegraphics[width=0.53\textwidth]{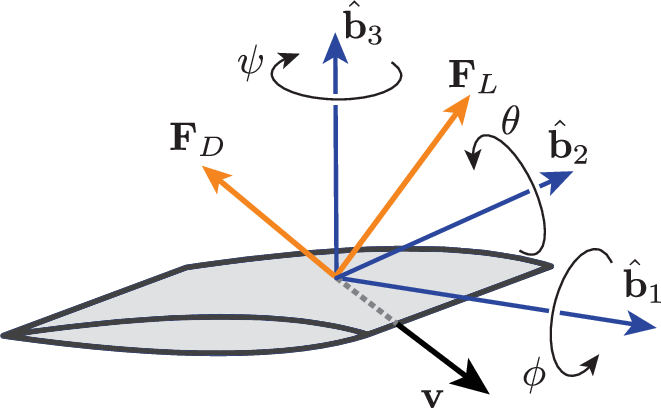} &
\includegraphics[width=0.37\textwidth]{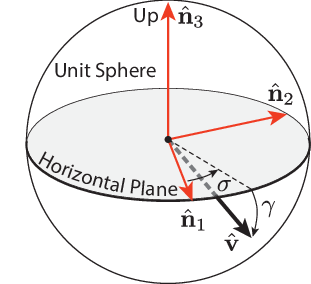}\\
{(a)} & {(b)}
\end{tabular}
\caption{(a) Body-fixed frame illustrating pitch, roll, and yaw angles. (b) Unit velocity vector and associated glide and azimuthal angle definitions used in deriving the three-dimensional equations of motion.}
\label{Wing}
\end{figure}

Figure~\ref{Wing} illustrates the relevant geometric quantities. 
Panel (a) shows the body-fixed frame attached to the airfoil and the definitions of pitch, roll, and yaw.
We note that while our axes convention differs from the typical aircraft convention (e.g., in which the third axis would be down), our angle convention matches the usual convention (e.g., positive yaw to the right).
Panel (b) depicts the unit velocity vector in inertial space and its associated glide and azimuthal angles. 

The three-dimensional velocity space encodes the relationship between velocity, glide angles, and the resulting acceleration vector. 
Within this representation, equilibrium glide states correspond to points where the net acceleration vanishes. 
As in the two-dimensional setting, the equilibrium glide states may be stable or unstable (usually saddles), and these states and their associated stable and unstable manifolds act as the primary organizing structures of the velocity dynamics.
By extending the polar diagram to a fully 3D velocity space, we obtain a unified geometric framework that captures the combined effects of lift, drag, gravity, and spatial orientation on the glider's motion.

\subsection{Equations of Motion in Cartesian Coordinates}
We write the inertial velocity $\vv$ of the wing in terms of its magnitude and direction,
\begin{equation}
    \vv = v \hat{\vv},
\end{equation}
where $\hat{\vv}$ is the unit vector in the velocity direction and $v$ is the velocity magnitude. This unit vector, in inertial frame components, is,
\begin{equation}
    \hat{\vv} = v_1 \hat{\nn}_1 + v_2 \hat{\nn}_2 + v_3 \hat{\nn}_3,
\end{equation}
thus, $v = \sqrt{v_1^2 + v_2^2 + v_3^2}$.
The velocity unit can be written in terms of the glide slope angle $\gamma$ and an azimuthal angle $\sigma$ as, 
\begin{equation}\label{velocity_angles}
    \hat{\vv} =   \underbrace{\cos{\gamma} \cos{\sigma}}_{v_1/v} \hat{\nn}_1 + \underbrace{\cos{\gamma} \sin{\sigma}}_{v_2/v} \hat{\nn}_2~\underbrace{- \sin{\gamma}}_{v_3/v} \hat{\nn}_3.
\end{equation}
where, referring to Figure \ref{Wing}, these angles are related to the inertial velocity components,
\begin{equation}\label{glide_and_azimuthal_angles}
    \gamma =  -\tan^{-1} \left( \frac{v_3}{\sqrt{v_1^2 + v_2^2}} \right), 
    \quad
    \sigma = \tan^{-1} \left( \frac{v_2}{v_1} \right),
\end{equation}

The equation of motion from Newton's Second Law is,
\begin{equation}\label{eom}
    m \dot{\vv}  = \mathbf{F}_D + \mathbf{F}_L - m \mathbf{g},
\end{equation}
which can be written in terms of the cartesian components ($v_1,v_2,v_3$) or spherical coordinates ($v,\gamma,\sigma$).
Note,  $m$ is the airfoil mass and $\mathbf{g} = g \hat{\nn}_3$, where $g$ is the acceleration due to gravity.

The relationship between the body $\mathcal{B}$-frame unit vectors $\{ \hat{\bbb}_1, \hat{\bbb}_2, \hat{\bbb}_3 \}$ and the  inertial $\mathcal{N}$-frame unit vectors $\{ \hat{\nn}_1, \hat{\nn}_2, \hat{\nn}_3 \}$ is encapsulated in the rotation matrix (written in the convention of \cite{Schaub2009}),
\begin{equation} \label{BN}
\footnotesize
[BN] =
\begin{bmatrix}
\cos\psi \cos\theta & -\sin\psi \cos\theta & \sin\theta \\
 -\cos\psi \sin\theta \sin\phi + \cos\phi \sin\psi &
 \sin\psi \sin\theta \sin\phi + \cos\phi \cos\psi &
\cos\theta \sin\phi \\
-\cos\phi \sin\theta \cos\psi - \sin\phi \sin\psi &
\sin\psi \sin\theta \cos\phi - \sin\phi \cos\psi &
\cos\theta \cos\phi
\end{bmatrix},
\end{equation}
where the usual aircraft convention is maintained: 
positive yaw   ($\psi$) and roll ($\phi$)  to the right, 
and positive pitch ($\theta$) is upward.

The drag force is $\mathbf{F}_D = -F_D \hat{\vv}$ while the lift force is taken as, 
\begin{equation}
    \mathbf{F}_L = F_L \hat{\vv} \times \hat{\bbb}_2,
\end{equation}
which ensures the lift force is orthogonal to both the direction of motion (and drag force) and the lateral airfoil axis. 
This yields a lift vector that points roughly upwards in the wind frame.
Writing the instantaneous airfoil lateral direction in the inertial
direction as $\hat{\bbb}_2 = b_{2x} \hat{\nn}_1 + b_{2y} \hat{\nn}_2 + b_{2z} \hat{\nn}_3$,
where each $b_{2i}$ can be written in terms of Euler angles $\boldsymbol{\theta} =(\psi,\theta,\phi)$ from the middle row of $[BN]$,
the lift force inertial direction is,
\begin{equation}
\hat{\vv} \times \hat{\bbb}_2 
= -  \hat{\bbb}_2  \times \hat{\vv} 
= -  \tilde{\bbb}_2 \hat{\vv},
\end{equation}
where the `tilde' notation \cite{Schaub2009} allows us to write the cross-product as a matrix product,
\begin{equation}\label{lateral_axis}
\tilde{\bbb}_2
= 
\begin{bmatrix}
    0 & -b_{2z} & b_{2y} \\
b_{2z} & 0 & -b_{2x} \\
- b_{2y} & b_{2x} &  0
\end{bmatrix}.
\end{equation}
To make the connection with the lift force clearer, we will rename $\tilde{\bbb}_2$ as $\tilde{\mathbf{L}}$, and thus,
\begin{equation}
\tilde{\mathbf{L}}(\boldsymbol{\theta}) =
\begin{bmatrix}
~~0 & -C & ~~B\\
~~C & ~~0  & -A~\\
-B~ & ~~A & ~~0
\end{bmatrix},
\end{equation}
where,
\begin{equation}
\label{ABC_full}
    \begin{split}
        A &= \cos \phi \sin \psi   -\cos \psi \sin \theta \sin \phi , \\
        B &= \cos \phi \cos \psi + \sin \psi \sin \theta \sin \phi , \\
        C &=  \cos \theta \sin \phi.
    \end{split}
\end{equation}

The form of the lift and drag force are the dynamic pressure times a lift and drag coefficient, respectively,
\begin{equation}
    F_L = \frac{\rho v^2}{2S} C_L(\alpha), \quad \
    F_D = \frac{\rho v^2}{2S} C_D(\alpha),
\end{equation}
where $\rho$ is the surrounding fluid density, $S$ is the airfoil effective surface area, and $\alpha=\tan^{-1}( w / u )$ is the airfoil angle of attack, 
\begin{equation}\label{alpha_full}
\footnotesize
\alpha = \tan^{-1} \left( 
\frac{
\cos\gamma \cos\phi \sin\theta \cos(\psi + \sigma) 
+ \cos\gamma \sin\phi \sin(\psi + \sigma) 
+ \sin\gamma \cos\phi \cos\theta
}{
\cos\gamma \cos\theta \cos(\psi + \sigma) 
- \sin\gamma \sin\theta
}
\right).
\end{equation} 
Note that angle of attack depends on the instantaneous airfoil velocity through the glide and azimuthal angles via \eqref{glide_and_azimuthal_angles}.

The equations of motion \eqref{eom} can be written as,
\begin{equation}
    m \dot{\vv} = -v C_D(\alpha(\vv)) \vv - v C_L(\alpha(\vv)) \tilde{\bbb}_2 \vv - m \mathbf{g},
\end{equation}
which is three scalar coupled, nonlinear ordinary differential equations for the evolution of $\vv = (v_1,v_2,v_3)^T$ 
which depends on several parameters: Euler angles $(\psi,\theta,\phi)$, physical parameters $(\rho,S,m)$, and lift and drag curves $(C_L,C_D)$. The last two are functions, thus, infinite-dimensional.

To simplify analysis, we non-dimensionalize time $(t)$, position component $(x_i)$, and velocity component $(v_i)$, respectively, using the  airfoil chord length $c$ and gravitational acceleration $g$,
\begin{equation}
    \tilde{t} = \frac{t}{\sqrt{c/g}}, \quad \tilde{x}_i = \frac{x_i}{c}, \quad \tilde{v}_i = \frac{v_i}{\sqrt{c g}}.
\end{equation}
In place of the three physical parameters $(\rho,S,m)$, non-dimensionalizing
leaves us with equations of motion that contain a single dimensionless parameter, the universal glide parameter \cite{yeaton2017global},
\begin{equation}
    \epsilon = \frac{\rho c S}{2 m}.
\end{equation}
For brevity, we do not show this derivation (see \cite{yeaton2017global,nave2019global}).

We can further rescale time and velocity in the equations using the glide parameter $\epsilon$ so it does not appear in the dynamic equations,  
\begin{equation}
    \bar{v}_i = \sqrt{\epsilon} \, \tilde{v}_i, \quad \bar{t} = \sqrt{\epsilon} \, \tilde{t},
\end{equation}
We are left with the re-scaled, non-dimensionalized 3D glider dynamic equations for $(\bar{v}_1,\bar{v}_2,\bar{v}_3)$,
\begin{equation}
\begin{split}
\Bar{v}_1^{\prime} &= 
- \Bar{v} C_D(\alpha) \Bar{v}_1
- \Bar{v} C_L(\alpha) \left( -\Bar{v}_2 C + \Bar{v}_3 B  \right)  \\
\Bar{v}_2^{\prime} &= 
- \Bar{v} C_D(\alpha) \Bar{v}_2
- \Bar{v} C_L(\alpha) \left( ~~\Bar{v}_1 C - \Bar{v}_3 A  \right) \\
\Bar{v}_3^{\prime} &= 
- \Bar{v} C_D(\alpha) \Bar{v}_3
- \Bar{v} C_L(\alpha) \left( -\Bar{v}_1 B + \Bar{v}_2 A
 \right)  - 1,
\end{split}
\end{equation}
where 
the prime notation denotes the derivative with respect to the re-scaled, non-dimensional time, $\frac{\rm d}{{\rm d} \bar{t}}$. 
For simplicity of exposition, we will drop the over bars and use the over-dot for the time derivative.
The remaining parameters are the orientation parameters $(\psi,\theta,\phi)$ and the lift and drag curves, viewed as infinite-dimensional functional parameters.
We note that the non-dimensionalized and re-scaled equations can be written in a compact form for $\vv = (v_1,v_2,v_3)^T$,
\begin{equation}\label{eom_compact}
\dot{\vv} = \mathbf{A}(\vv) \vv - \hat{\nn}_3,
\end{equation}
where,
\begin{equation}\label{eom_matrix}
\mathbf{A}(\vv) = \mathbf{D}(\vv) + \mathbf{L}(\vv),
\quad 
\hat{\nn}_3 
=
\begin{bmatrix}
0 \\
0 \\
1
\end{bmatrix},
\end{equation}
where we note that $3\times3$ matrix $\mathbf{A}(\vv)$ can be decomposed into the effect of drag,
$\mathbf{D}(\vv) = - v C_D \mathbf{I}_3$, proportional to the identity matrix, and the effect of lift, represented by $\mathbf{L}(\vv) = -\mathbf{L}(\vv)^T$, a skew-symmetric matrix,
\begin{equation}\label{eom_matrix}
\mathbf{L}(\vv) = -v C_L
\tilde{\mathbf{L}}, \quad
{\rm where}
\quad
\tilde{\mathbf{L}} = \tilde{\bbb}_2, 
\end{equation}
Thus, $\tilde{\mathbf{L}}$ depends on the Euler angles, $\boldsymbol{\theta}=(\psi,\theta,\phi)$, through the orientation of the airfoil lateral axis via \eqref{lateral_axis}.

\paragraph{Jacobian.}
The compact form makes the calculation of the Jacobian particularly simple, since the identity $\mathbf{I}_3$ and the skew-symmetric matrix $\tilde{\mathbf{L}}$ are independent of $\vv$. 
All dependence on $\vv$ is contained in the terms $v = \sqrt{v_1^2 + v_2^2 +v_3^2}$,
$C_D = C_D(\alpha(\vv))$, and $C_L = C_L(\alpha(\vv))$.
We write $\nabla(v C_{D})$ for the gradient of $v C_{D}(\alpha(\vv))$ with respect to $(v_{1},v_{2},v_{3})$, and similarly $\nabla(v C_{L})$.
The Jacobian of \eqref{eom_compact}
at a point $\vv$ is,
\begin{equation}\label{jacobian_cart}
\mathbf{J}(\vv) = \mathbf{A}(\vv)
-\left( \nabla(v C_{D})\cdot \vv \right) \mathbf{I}_3
-\left( \nabla(v C_{L})\cdot \vv \right) \tilde{\mathbf{L}}.
\end{equation}

\subsection{Spherical Coordinate Formulation}

To gain geometric insight into the glider's motion, we next express the body-fixed velocity vector in wind-aligned spherical coordinates. Using the angles defined in \eqref{velocity_angles}, we have for $ \vv_p = (v,\gamma,\sigma)^T$,
\[
v \;=\;\sqrt{v_1^2 + v_2^2 + v_3^2},\quad
\gamma \;=\;-\,\tan^{-1}\!\Bigl(\frac{v_3}{\sqrt{v_1^2+v_2^2}}\Bigr),\quad
\sigma \;=\;\tan^{-1}\!\Bigl(\frac{v_2}{v_1}\Bigr),
\]
we have
\[
\begin{bmatrix}v_1\\v_2\\v_3\end{bmatrix}
=
v\,
\underbrace{
\begin{bmatrix}
\cos\gamma\,\cos\sigma\\
\cos\gamma\,\sin\sigma\\
-\sin\gamma
\end{bmatrix}
}_{\hat{\vv}}.
\]
Taking the time derivative, we get,
\begin{equation}\label{derivative_relationship}
    \underbrace{
\begin{bmatrix}
\dot v_1\\
\dot v_2\\
\dot v_3
\end{bmatrix}
}_{\dot{\mathbf v}}
=
\underbrace{
\begin{bmatrix}
\cos\gamma\cos\sigma & -v \sin\gamma\cos\sigma & -v \cos\gamma\sin\sigma\\
\cos\gamma\sin\sigma & -v \sin\gamma\sin\sigma & ~~ v \cos\gamma\cos\sigma\\
-\sin\gamma          & -v\cos\gamma            &  0
\end{bmatrix}
}_{\mathbf{M}(\vv_p)}
\underbrace{\begin{bmatrix}
\dot v\\
\dot\gamma\\
\dot\sigma
\end{bmatrix}
}_{\dot{\vv}_p}.
\end{equation}
Now as long as
$\det \left(\mathbf{M}(\vv_p)\right) = v\cos\gamma \neq 0$,
we have the inverse,
\begin{equation}
\mathbf{M}(\vv_p)^{-1} =
\begin{bmatrix}
\cos \gamma \cos \sigma &
\cos \gamma \sin \sigma &
 -\sin \gamma \\
 -\frac{\sin \gamma \cos \sigma}{v} &
 -\frac{\sin \gamma \sin \sigma}{v}&
 -\frac{\cos \gamma}{v} \\
 -\frac{\sin \sigma}{v \cos \gamma} &
  \frac{\cos \sigma}{v \cos \gamma} &
0
\end{bmatrix}.
\end{equation}
From \eqref{eom_compact} and \eqref{derivative_relationship}, we get,
\begin{equation}
    \dot{\vv}_p = \mathbf{M}(\vv_p)^{-1}\left(\mathbf{A}(\vv) \vv - \hat{\nn}_3\right),
\end{equation}
And after some algebra, one finds,
\begin{equation}
\label{spherical_vector}
    \dot{\vv}_p 
    =-v^2  C_D(\alpha) \mathbf{M}(\vv_p)^{-1}\hat{\vv}
     -v^2  C_L(\alpha) \mathbf{M}(\vv_p)^{-1}\tilde{\mathbf{L}}(\boldsymbol{\theta}) \hat{\vv}
     - \mathbf{M}(\vv_p)^{-1}\hat{\nn}_3.
\end{equation}

Evaluating the vector contribution of drag and gravity, we obtain,
\begin{equation}
\mathbf{M}(\vv_p)^{-1} \hat{\vv} = \begin{bmatrix}
1 \\
0 \\
0
\end{bmatrix}, \quad {\rm and} \quad
\mathbf{M}(\vv_p)^{-1}\hat{\nn}_3 = \begin{bmatrix}
-\sin \gamma \\
-\frac{\cos \gamma}{v} \\
0
\end{bmatrix},
\end{equation}
and we also note,
\begin{equation}
\tilde{\mathbf L}(\boldsymbol{\theta}) \hat{\vv}
=
\begin{bmatrix}
  -C\,\cos\gamma\,\sin\sigma\;-\;B\,\sin\gamma\\[0pt]
~\;C\,\cos\gamma\,\cos\sigma\;+\;A\,\sin\gamma\\[0pt]
\cos\gamma\,\bigl(-B\,\cos\sigma + A\,\sin\sigma\bigr)
\end{bmatrix}.
\end{equation}
For the vector contribution of lift, we have,
\begin{equation}
\mathbf{M}(\vv_p)^{-1}\tilde{\mathbf{L}}(\boldsymbol{\theta}) \hat{\vv} =
\begin{bmatrix}
 0
\\
-\frac1v\bigl(
  - B\,\cos \sigma +  A\,\sin\sigma \bigr)
\\
-\frac1{v\,\cos\gamma}\bigl(
- C \cos \gamma- \sin \gamma (A\,\cos\sigma + B\,\sin\sigma ) \bigr)
\end{bmatrix}.
\end{equation}

The equations of motion in spherical coordinates \eqref{spherical_vector} are then,
\begin{equation}\label{eom_spherical_full}
    \begin{split}
 \dot v &= -v^2 C_D(\alpha)  +\sin \gamma ,\\
\dot \gamma &= v C_L(\alpha) \bigl(
- B\,\cos\sigma +  A\,\sin\sigma
\bigr) + \frac{\cos \gamma}{v},\\
\dot \sigma &= \frac{v C_L(\alpha)}{\cos\gamma}\bigl(
 - C \cos \gamma- \sin \gamma (A\,\cos\sigma + B\,\sin\sigma )
\bigr).
    \end{split}
\end{equation}

\paragraph{Small angle approximation.}
If we assume small yaw, roll, and azimuthal angles---keeping terms through second order in these angles---we get an approximate expression for the angle of attack $\alpha$ from \eqref{alpha_full} which involves all five angles, 
\begin{equation}\label{angle_of_attack}
\alpha \approx  \theta + \gamma + \phi (\psi + \sigma),
\end{equation}
where we have the longitudinal version ($\alpha = \theta + \gamma$) corrected by the second order term $\phi (\psi + \sigma)$, and terms third order and higher in the expression are neglected.

In $\tilde{\mathbf L}(\boldsymbol{\theta})$, if we keep only terms through second order in the Euler angles $\boldsymbol{\theta}$, we have an approximate expression for $A$, $B$, and $C$, via \eqref{ABC_full}, 
\begin{equation}
\label{ABC_approx}
    \begin{split}
        A &\approx \psi - \theta \phi, \\
        B &\approx 1 - \tfrac{1}{2}(\phi^2 + \psi^2), \\
        C &\approx \phi.
    \end{split}
\end{equation}
Notice that even for a small roll and yaw approximation, all factors in the lift depend on roll.
This yields the following approximate equations of motion,
\begin{equation}
\small
    \begin{split}
 \dot v &\approx -v^2 C_D(\alpha)  +\sin \gamma ,\\
\dot \gamma &\approx v C_L(\alpha) \bigl(
- (1 - \tfrac{1}{2}(\phi^2 + \psi^2))\,\cos\sigma +  (\psi - \theta \phi)\,\sin\sigma
\bigr) + \frac{\cos \gamma}{v},\\
\dot \sigma &\approx \frac{v C_L(\alpha)}{\cos\gamma}\bigl(
 - \phi \cos \gamma- \sin \gamma ((\psi - \theta \phi)\,\cos\sigma + (1 - \tfrac{1}{2}(\phi^2 + \psi^2))\,\sin\sigma )
\bigr).
    \end{split}
\end{equation}

\subsection{Lift and Drag as Functional Parameters}

The behavior of the three-dimensional glider model is determined by the body orientation, specified by pitch, roll, and yaw angles, together with the aerodynamic force laws. 
The aerodynamic coefficients $C_L$ and $C_D$ are functions of $\alpha$, encapsulating the shape-dependent aerodynamic response of each airfoil. Formally, $\alpha$ is along the circle, $S^1$, and thus,
\[
C_L: S^1 \rightarrow I_L \subset \mathbb{R}, \qquad 
C_D: S^1 \rightarrow I_D \subset \mathbb{R}^+,
\]
with $C_L$ allowed to change sign and $C_D$ strictly positive to ensure non-negative drag. These mappings govern the lift and drag forces and, through them, the geometry of the phase-space structures that organize the glide dynamics.

We apply this framework to three representative airfoils spanning both biological and engineered gliders. The first is the cross-sectional profile of the flying snake \textit{Chrysopelea paradisi}, whose bluff geometry generates substantial lift through complex unsteady aerodynamics. The second is the canonical NACA 0012, a symmetric airfoil widely studied in aerodynamics and often used as a benchmark. The third is the Zimmerman airfoil, a low-Reynolds-number planform whose outline closely matches the patagial extension of \textit{Draco} lizards. Each airfoil is characterized by its own pair of coefficient functions $C_L(\alpha)$ and $C_D(\alpha)$, obtained from experimental data and fitted to smooth analytic forms for incorporation into the model. These functions directly determine the shape of the terminal velocity manifold, the nature of trajectories converging to it, and the organization of equilibrium points across pitch, roll, and yaw. The unified 3D framework therefore provides a direct way to compare biological and engineered shapes, linking aerodynamic properties to phase-space geometry, stability, and glide efficiency.

\begin{figure}[h!]
  \centering
  \vspace{- 0 cm}
  \begin{subfigure}[b]{0.5\textwidth}
    \centering
    \includegraphics[width=\linewidth]{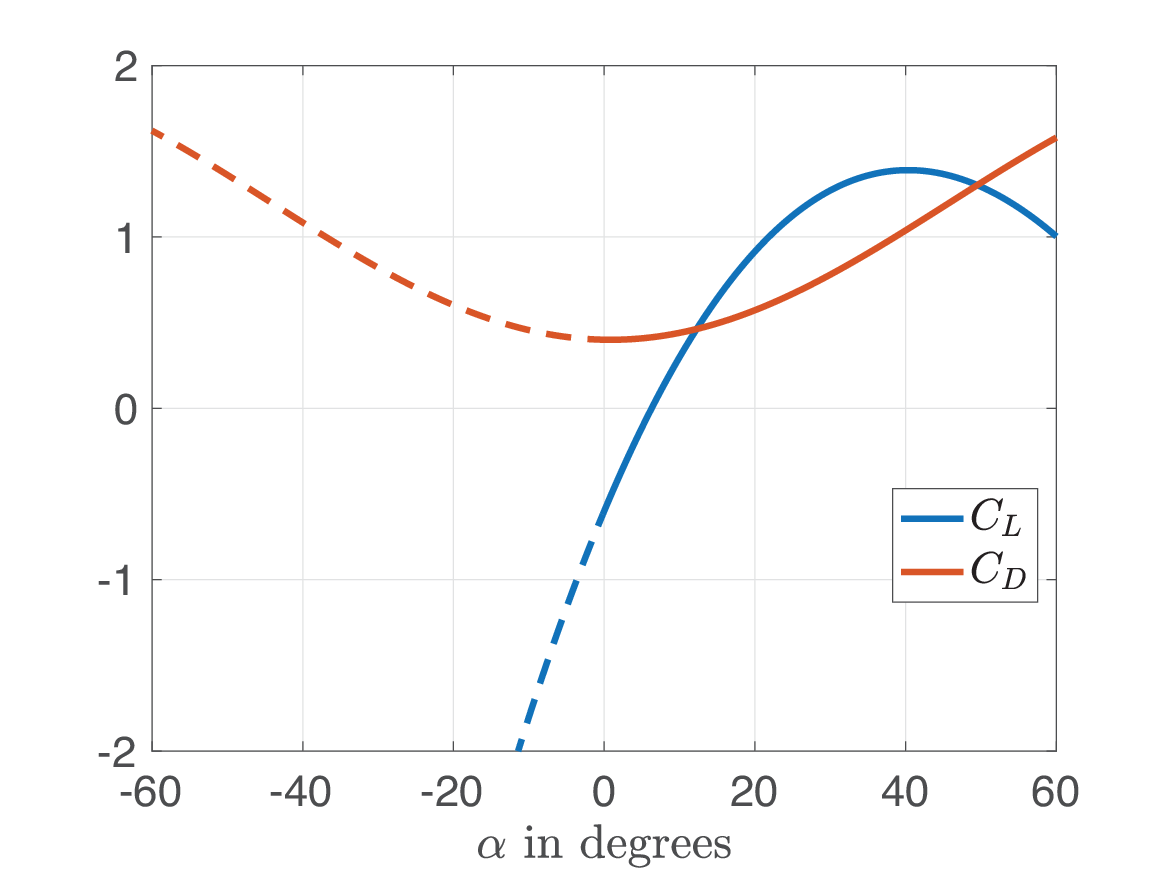}
    \caption{}
    \label{fig:CL_CD_idealized}
  \end{subfigure}
  \hfill
  \begin{subfigure}[b]{0.3\textwidth}
    \centering
       \raisebox{1cm}{
    \includegraphics[width=\linewidth]{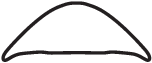} }
    \caption{}
    \label{fig:airfoil_idealized}
  \end{subfigure}

  \begin{subfigure}[b]{0.5\textwidth}
    \centering
    \includegraphics[width=\linewidth]{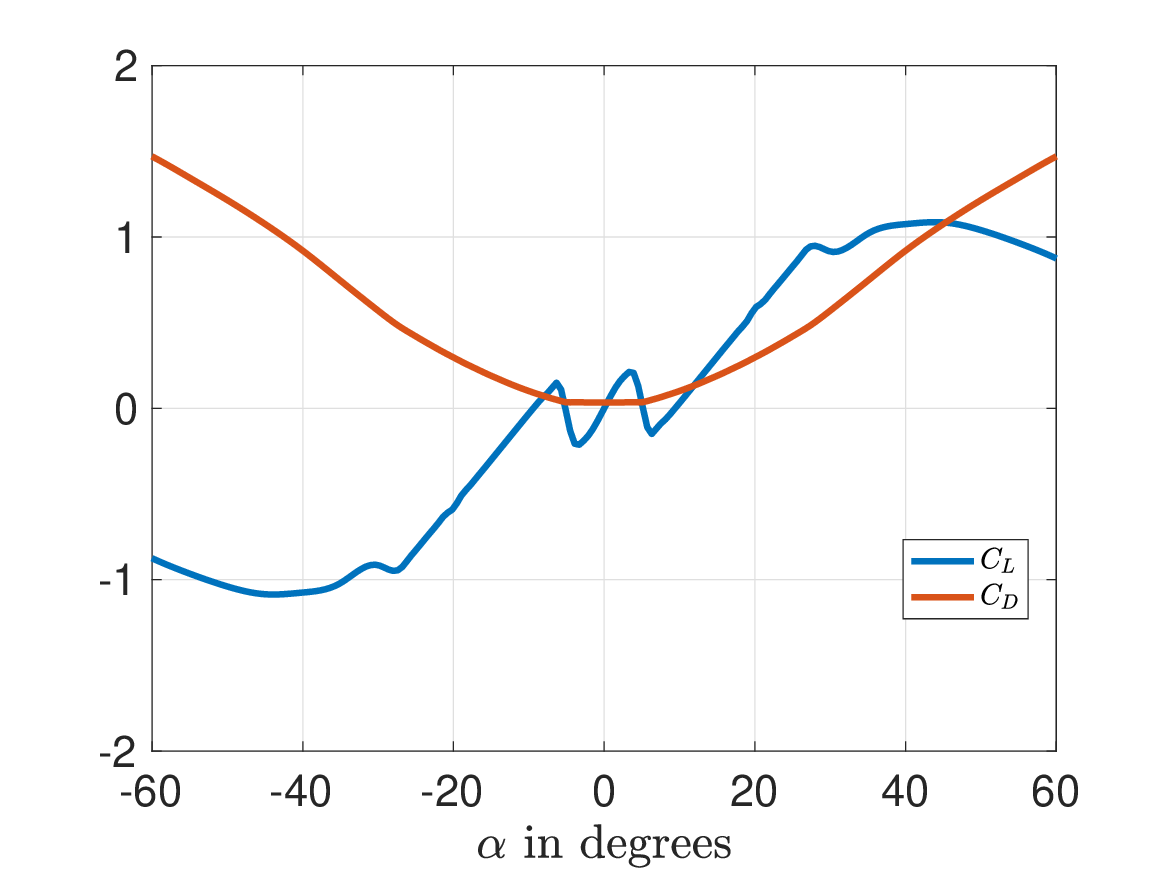}
    \caption{}
    \label{fig:CL_CD_NACA}
  \end{subfigure}
  \hfill
  \begin{subfigure}[b]{0.3\textwidth}
    \centering
    \raisebox{1.4cm}{
    \includegraphics[width=\linewidth]{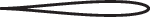} }
    \caption{}
    \label{fig:airfoil_NACA}
  \end{subfigure}

  \begin{subfigure}[b]{0.5\textwidth}
    \centering
    \includegraphics[width=\linewidth]{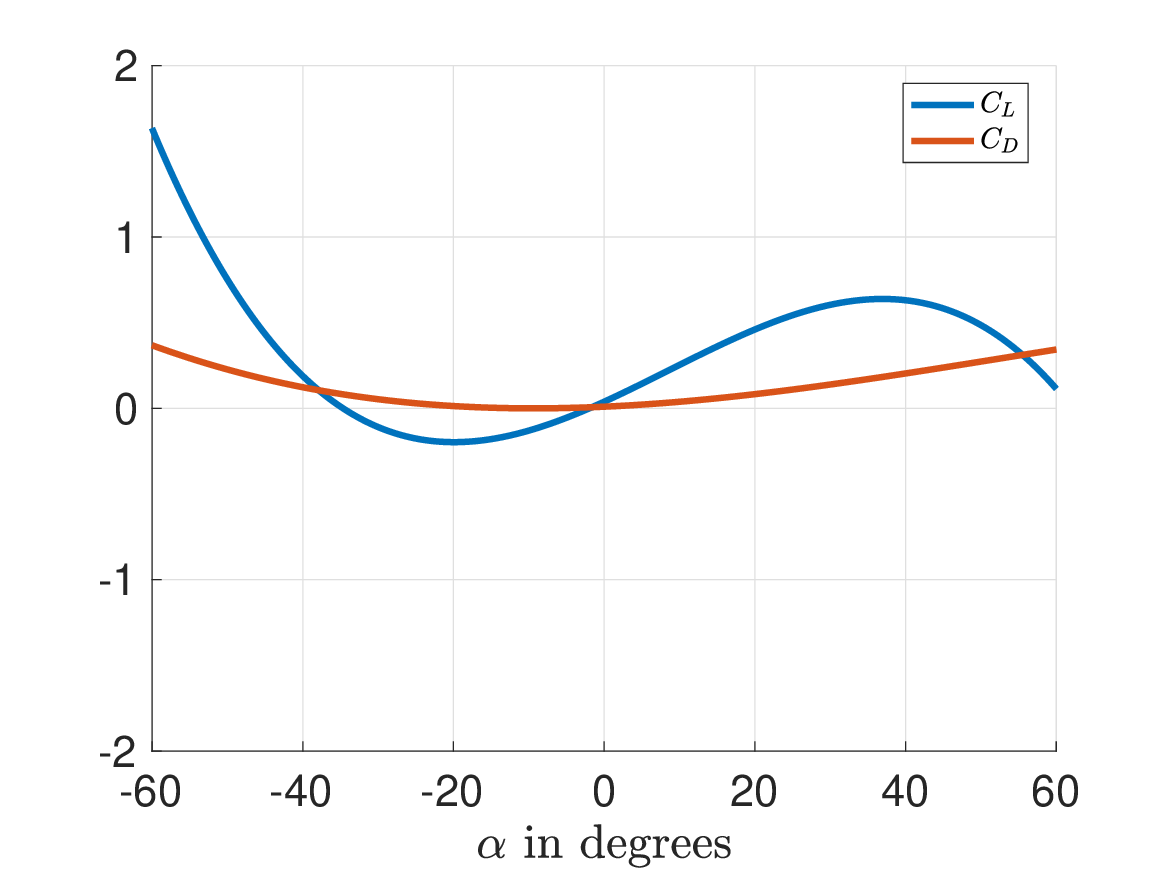}
    \caption{}
    \label{fig:CL_CD_zimm}
  \end{subfigure}
  \hfill
  \begin{subfigure}[b]{0.3\textwidth}
    \centering
    \includegraphics[width=\linewidth]{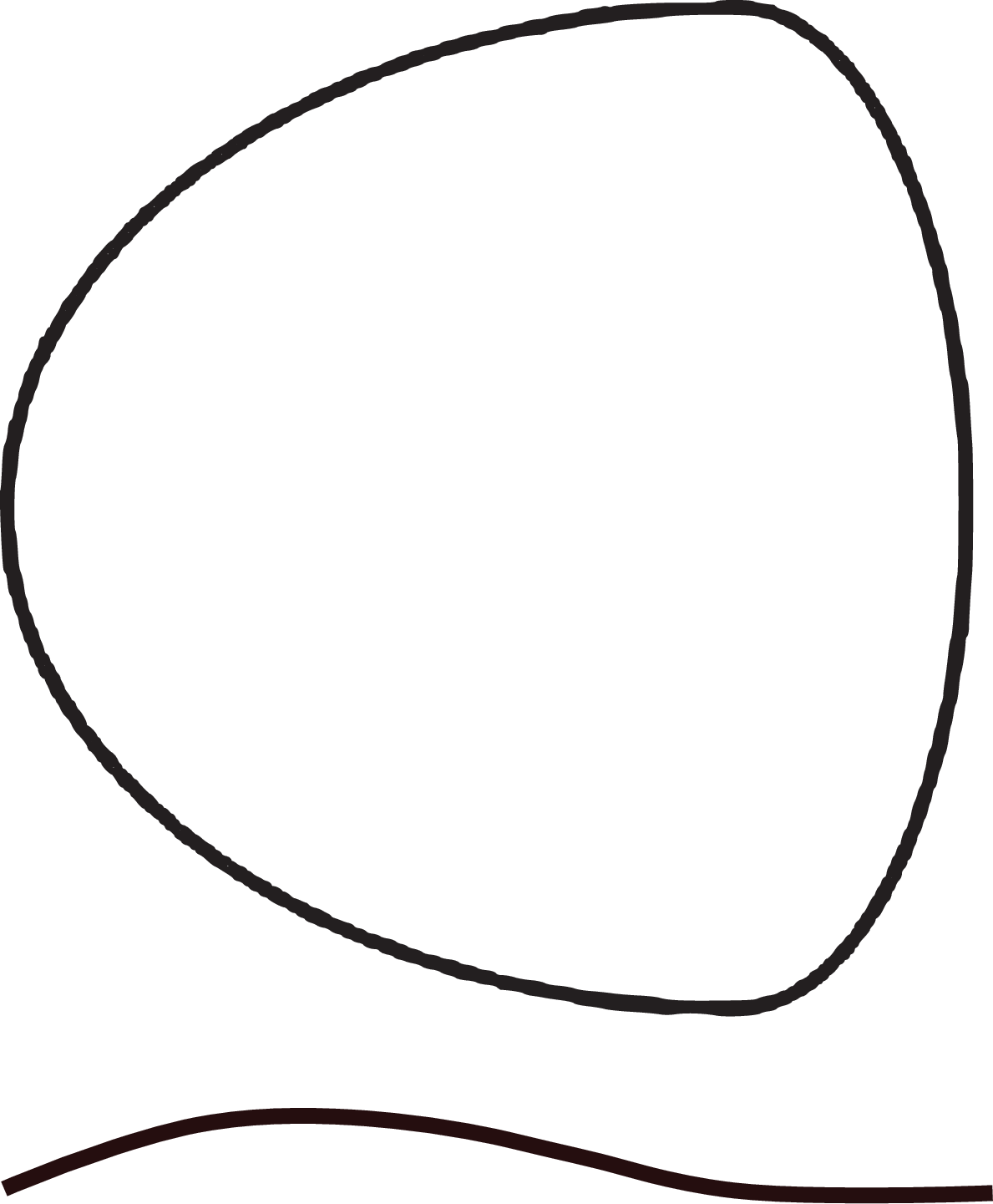} 
    \caption{}
    \label{fig:airfoil_zimm}
  \end{subfigure}

  \caption{%
    Left column: $C_L$ and $C_D$ curves for each airfoil.  
    Right column: corresponding airfoil planform shapes.  
    (a,b) Analytical Snake Airfoil. (c,d) NACA\,0012. (e,f) Zimmerman / Draco where both the top view and side view of the Zimmerman planform are shown. %
  }
  \label{fig:CL_CD_with_airfoils}
\end{figure}

\paragraph{Flying Snake, \textit{Chrysopelea paradisi}, Analytical Approximation.}
We begin with a biological example based on the flying snake \textit{Chrysopelea paradisi}, whose cross-sectional airfoil shape has been quantified previously \cite{holden2014aerodynamics, jafari2014theoretical, yeaton2020undulation}. 
During gliding, the snake laterally flattens its body to form a fore-aft symmetric bluff airfoil capable of generating significant lift.
The lift and drag coefficients, $C_L(\alpha)$ and $C_D(\alpha)$ in Figure \ref{fig:CL_CD_with_airfoils}, obtained from both computational study \cite{krishnan2014} and water channel experiments \cite{holden2014aerodynamics} spanning $-10^\circ \le \alpha \le 60^\circ$, are represented in our model through a smooth analytical approximation (shown as dashed curves), extended to $-60^\circ \le \alpha \le 60^\circ$ here.
This approximation to a wider range of angles of attack enables a three-dimensional construction of the terminal velocity manifold.
In extending the earlier 2D analyses \cite{yeaton2017global}, 
the 3D formulation highlights how pitch, roll, and yaw interact with the approximated aerodynamic response of the \textit{C.\ paradisi} airfoil to produce its characteristic gliding dynamics.

\paragraph{NACA 0012.}
As a canonical engineering example, we consider the NACA 0012 airfoil, a symmetric profile with extensively documented aerodynamic characteristics. 
Experimental measurements spanning a broad range of angles of attack (see \cite{abbott1949theory}) were used to generate smooth analytical approximations of the lift and drag curves, $C_L(\alpha)$ and $C_D(\alpha)$, as seen in Figure 2. 
These approximated functions were incorporated into the three-dimensional glider model to construct the corresponding TVM and examine its dependence on pitch, roll, and yaw. 
Owing to its symmetry and well-characterized aerodynamic response, the NACA 0012 serves as a useful baseline against which the behavior of biological airfoils can be compared, as in \cite{nave2019global}.

\paragraph{Flying \textit{Draco} Lizard, Approximated by Zimmerman Airfoil.}
As a second bio-inspired case study, we examine the Zimmerman airfoil, originally designed for low-speed aircraft and noted for its high lift-to-drag performance at low Reynolds numbers. 
Its planform closely resembles the rib-supported patagium of \textit{Draco} lizards, making it an effective engineered analogue of a biological glider. Aerodynamic measurements of $C_L(\alpha)$ and $C_D(\alpha)$ for this airfoil were smoothed using polynomial fitting to obtain analytical approximations suitable for use in the 3D model. 
The resulting TVM highlights how the Zimmerman's aerodynamic characteristics support efficient and robust gliding behavior. 
By analyzing trajectories as they converge to and evolve along this manifold, we relate its high $C_L/C_D$ ratio to its ability to maintain stable and easily accessible  glide states, consistent with observations of \textit{Draco} lizards in nature~\cite{khandelwal2022combined,khandelwal2023convergence}.

\section{Equilibrium Points and Bifurcation Analysis}

We begin our investigation of the three-dimensional velocity phase-space structure by identifying the equilibrium glide states of the system. 
We begin by characterizing how equilibrium glide states vary with body orientation. 
For prescribed Euler angles $\boldsymbol{\theta}$,
 equilibria are obtained when the right-hand side of the equations of motion, \eqref{eom_compact} or \eqref{eom_spherical_full}, vanishes; in other words, when the aerodynamic, gravitational, and inertial forces balance so that the velocity remains constant, producing a steady glide. 
These equilibria play a central role in shaping the surrounding phase space, and their qualitative changes under parameter variation form the basis of the bifurcation analysis that follows.

\paragraph{Equilibrium Points.}
The equilibrium conditions take their simplest form in the spherical coordinate formulation.
Setting the right-hand sides of \eqref{eom_spherical_full} to zero yields the relations,
\begin{equation}
\begin{split}\label{equilibrium}
    v^* &= \sqrt{ \frac{\sin \gamma^*}{C_D(\alpha^*)}}, \\
    \gamma^* &= \arccot \left( \frac{C_L}{C_D}(\alpha^*)\left( -A \sin \sigma^* + B \cos \sigma^* \right) \right),\\
    \sigma^* & = \arctan \left( \frac{C \frac{C_L}{C_D}(\alpha^*) + A}{C \frac{C_L}{C_D}(\alpha^*) - B} \right),
\end{split}
\end{equation}
where $\alpha^*$ is the equilibrium angle of attack and depends implicitly on the glide and heading angles, $\gamma^*$ and $\sigma^*$, through \eqref{alpha_full}. 
Because $\gamma^*$ and $\sigma^*$ are coupled, they must be solved for simultaneously.
Importantly, the location of the equilibrium points depends parametrically on the Euler angles, which serve as fixed orientation parameters in the model.

We seek to characterize how equilibrium glide states vary with body orientation.
For prescribed Euler angles $(\theta,\phi,\psi)$, equilibria are obtained by solving \eqref{equilibrium} for the coupled glide and heading angles $(\gamma^*,\sigma^*)$,
with equilibrium stability determined from the Jacobian \eqref{jacobian_cart}.
Before examining the more complex global bifurcation structure across airfoils, it is useful to first consider a planform whose equilibria vary smoothly and remain single-valued across orientation space. The Zimmerman airfoil provides such a baseline case.


\subsection{Orientation Dependence of Equilibria: The Zimmerman Airfoil}

\begin{figure}[H]
    \centering

    \begin{subfigure}[t]{0.32\textwidth}
    \includegraphics[width=\linewidth]{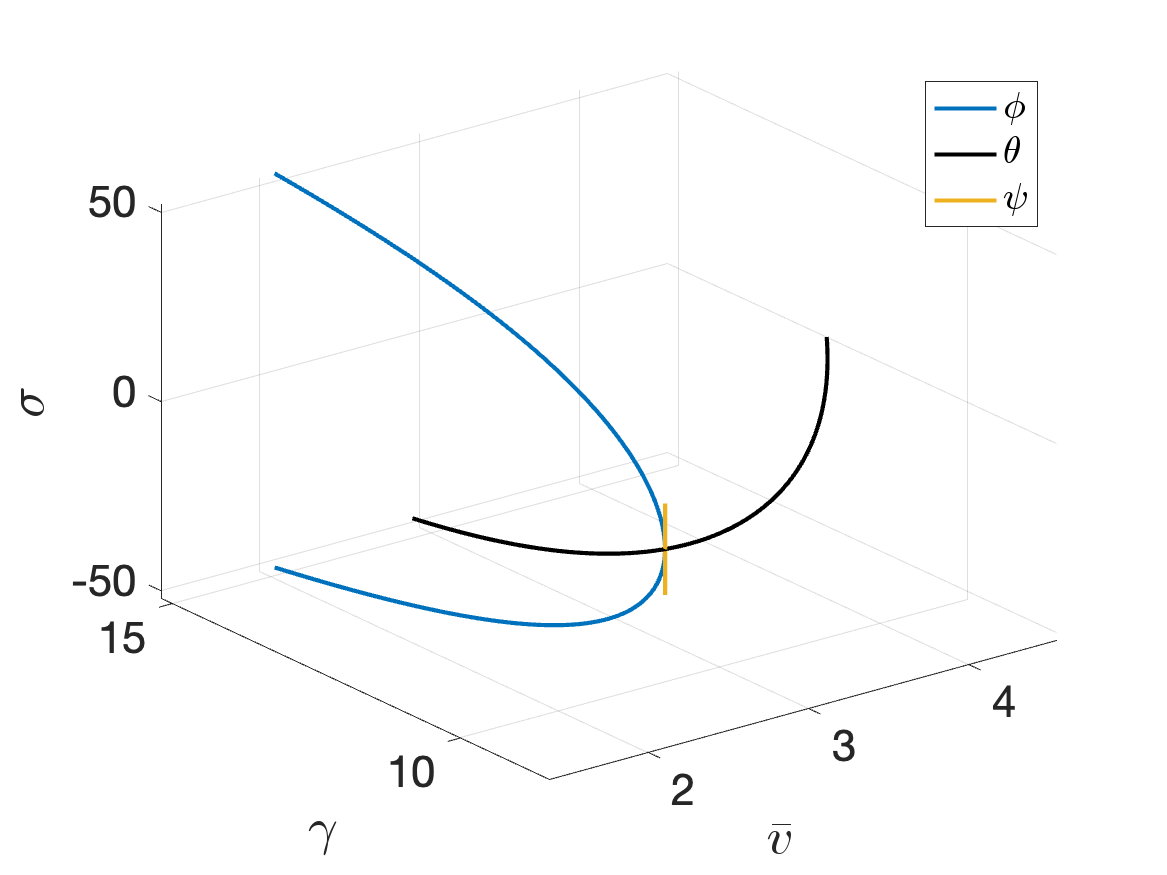}
    \caption{}
\end{subfigure}
\hfill
\begin{subfigure}[t]{0.32\textwidth}
    \includegraphics[width=\linewidth]{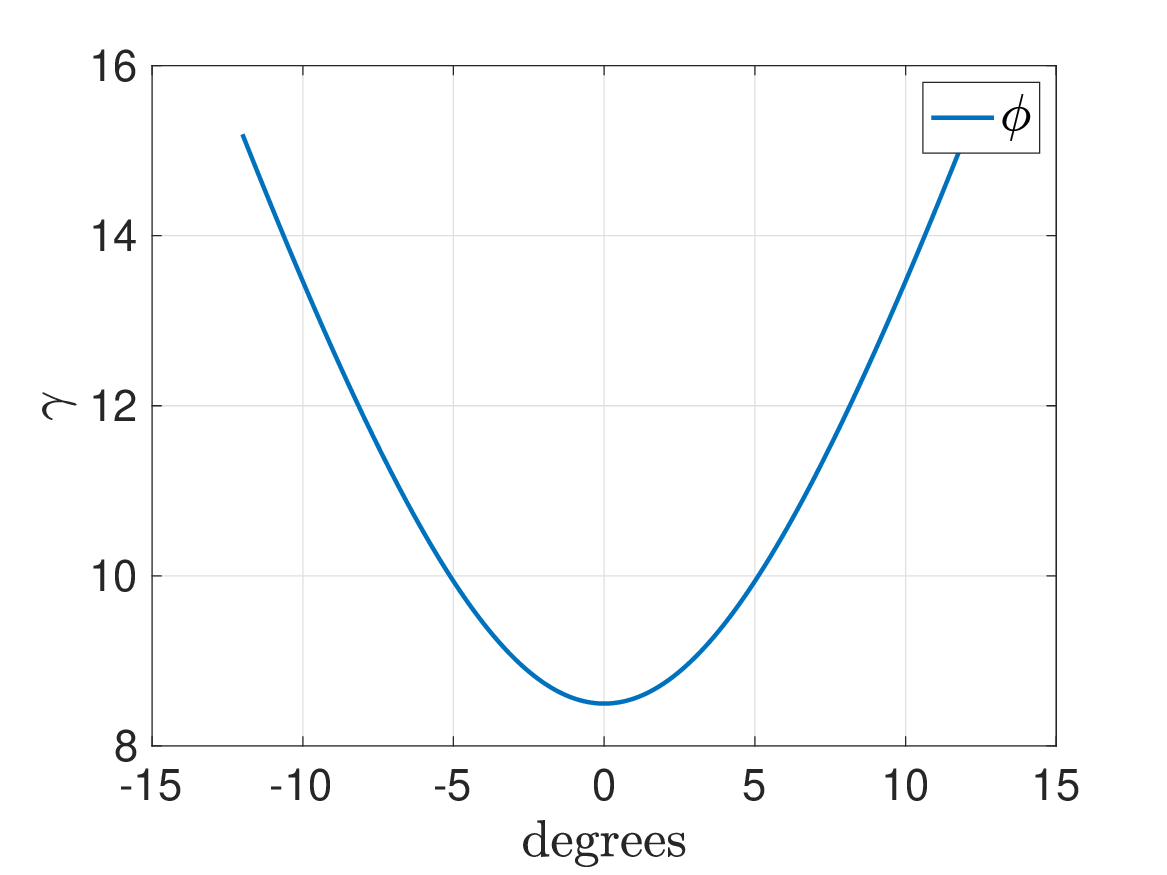}
    \caption{}
\end{subfigure}
\hfill
\begin{subfigure}[t]{0.32\textwidth}
    \includegraphics[width=\linewidth]{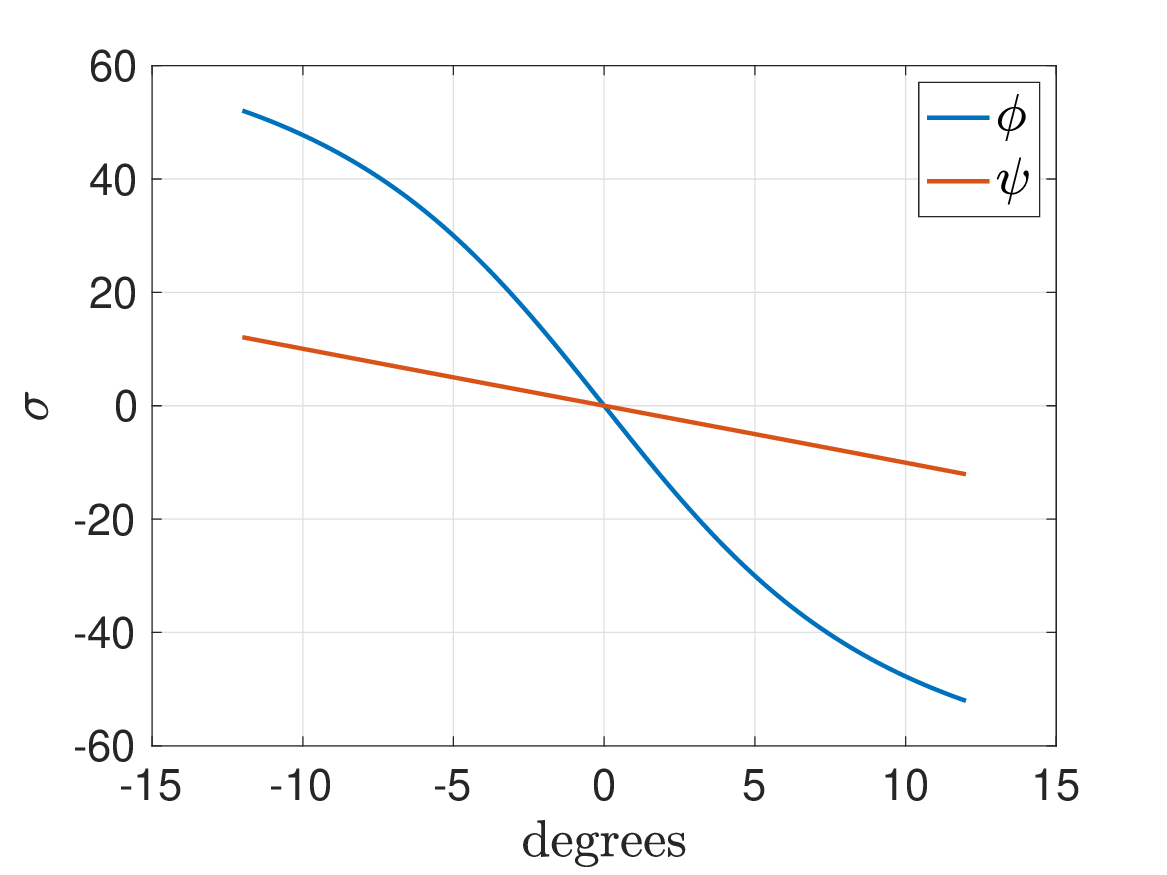}
    \caption{}
\end{subfigure}

\caption{
Variation of the Zimmerman airfoil equilibrium with body orientation parameters $(\phi,\theta,\psi)$.  
(a) The full equilibrium locus shown in $(\gamma^*,\sigma^*,v^*)$, illustrating how the steady glide state bends smoothly through velocity space as the Euler angles change.  
(b) Dependence of the equilibrium glide angle $\gamma^*$ on roll $\phi$, showing that increasing roll leads to steeper descent.  
(c) Sensitivity of the azimuthal angle $\sigma^*$ to roll and yaw. Roll has the dominant effect, indicating that $\phi$ is the primary steering parameter in this model.
}
  \label{fig:Zimmerman_eq}
\end{figure}

Figure~\ref{fig:Zimmerman_eq} illustrates how the shallow stable equilibrium glide state of the Zimmerman airfoil varies with changes in body orientation parameters $(\phi,\theta,\psi)$. 
The three-dimensional locus in $(\gamma^*,\sigma^*,v^*)$ is smooth and free of folds or self-intersections, reflecting the robustness of this planform. 
Panel (b) shows that the glide angle $\gamma^*$ attains its minimum at small roll $\phi$, yielding the shallowest descent path at near-symmetric alignment. 
As $|\phi|$ increases, $\gamma^*$ rises monotonically, indicating steeper glides.

The azimuthal equilibrium angle $\sigma^*$ responds far more strongly to roll than to yaw, as shown in panel (c). 
This reveals that roll is a significantly more effective steering input than yaw for the Zimmerman, consistent with its high $C_L/C_D$ ratio and lateral aerodynamic leverage.

Analytically, these variations can be approximated using small-angle expansions of \eqref{equilibrium}.
Up to second order in the angles, we have,
\begin{equation}
    \sigma^* = -\psi - \phi \frac{C_L}{C_D}(\alpha^*) + \phi \left[ \theta - \psi \frac{C_L}{C_D}(\alpha^*) - \phi \left( \frac{C_L}{C_D}(\alpha^*) \right)^2 \right],
\end{equation}
with the equilibrium angle of attack, from \eqref{angle_of_attack}, given approximately by,
\begin{equation}\label{angle_of_attack_equil}
\alpha^* \approx  \theta + \gamma^* + \phi (\psi + \sigma^*).
\end{equation}
Similarly, the equilibrium glide angle $\gamma^*$ up to second order is,
\begin{equation}
    \gamma^* = \frac{\pi}{2} + \frac{C_L}{C_D}(\alpha^*) \left(-1 + \sigma^*\psi + \frac{1}{2} 
    \left( \sigma^{*2} + \phi^2 + \psi^2 \right) \right),
\end{equation}
These expressions confirm analytically what is visible in Figure~\ref{fig:Zimmerman_eq}: roll couples strongly into the equilibrium direction of motion, while yaw has comparatively mild influence.

\subsection{Comparative Structure of Equilibria Across Airfoils}

\begin{figure}[!h]
  \centering

  \begin{subfigure}[t]{\textwidth}
    \centering
    \includegraphics[width=0.35\textwidth]{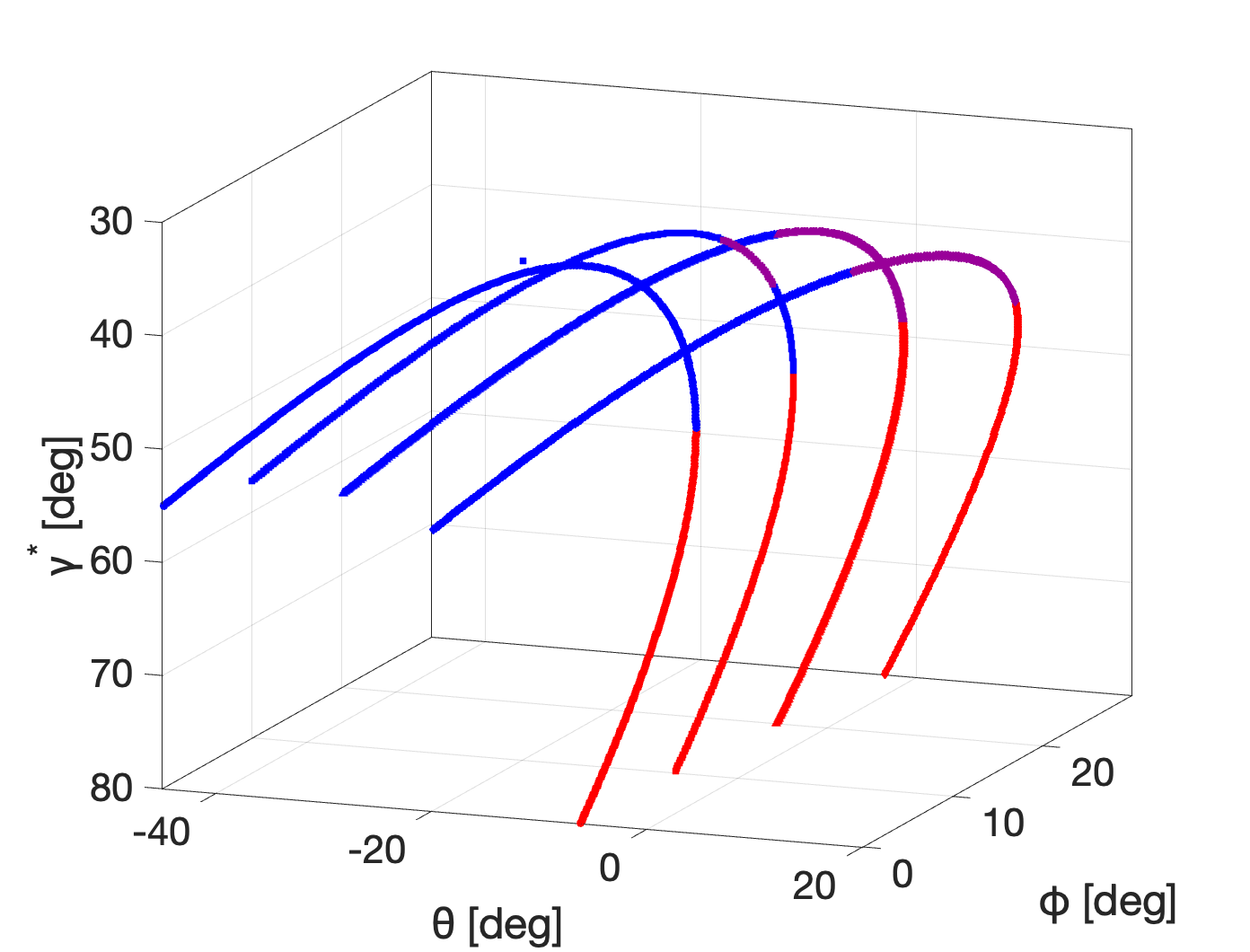}\hfill
    \includegraphics[width=0.35\textwidth]{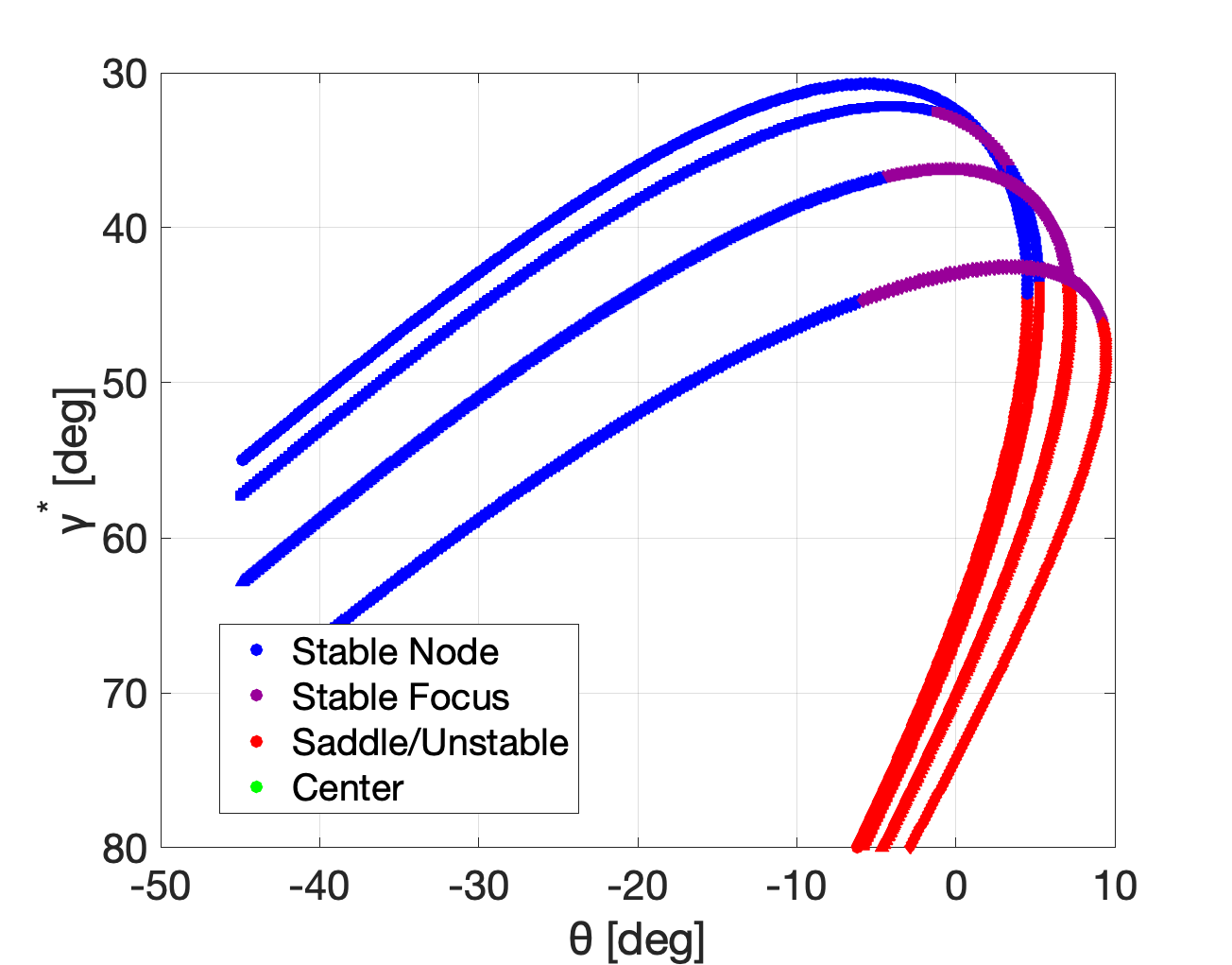}
    \caption{Snake model}
    \label{fig:idealized_filtered_overlay}
  \end{subfigure}

  \vspace{0em} 

  \begin{subfigure}[t]{\textwidth}
    \centering
    \includegraphics[width=0.35\textwidth]{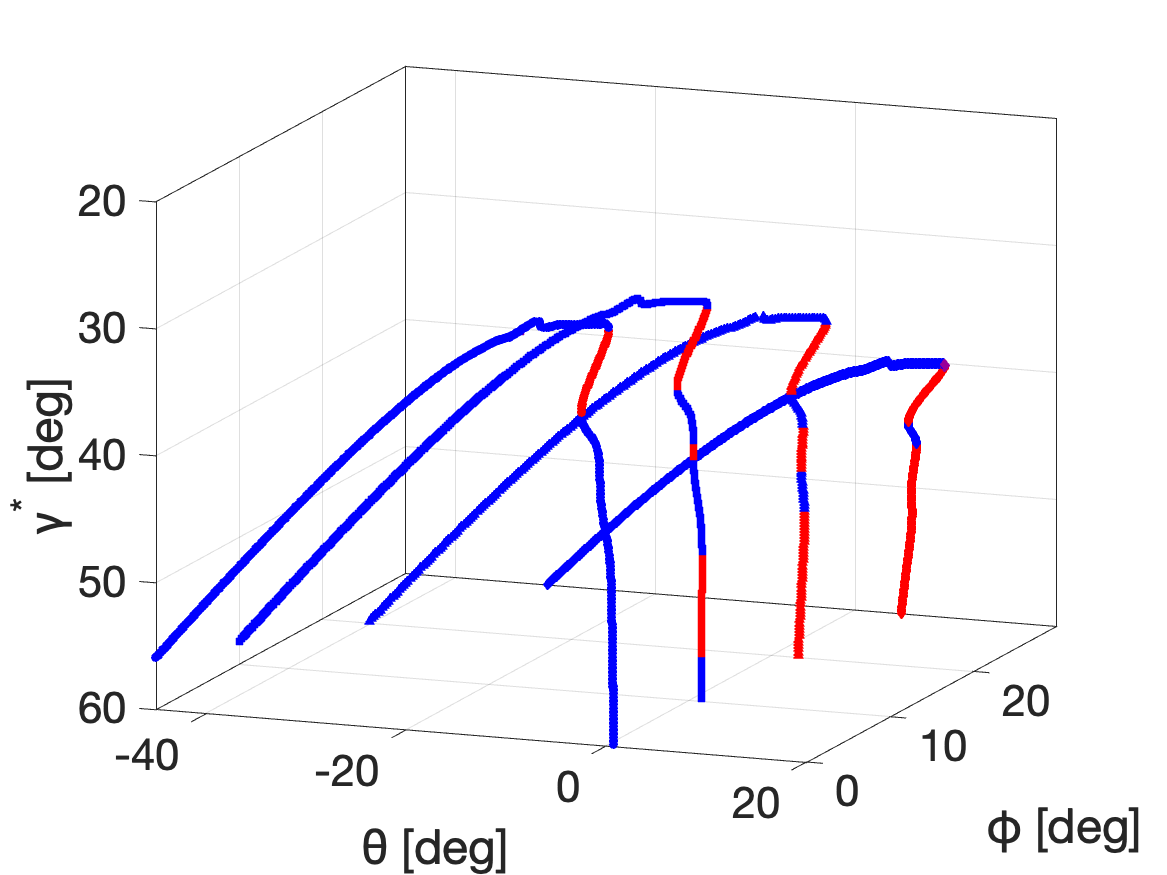}\hfill
    \includegraphics[width=0.35\textwidth]{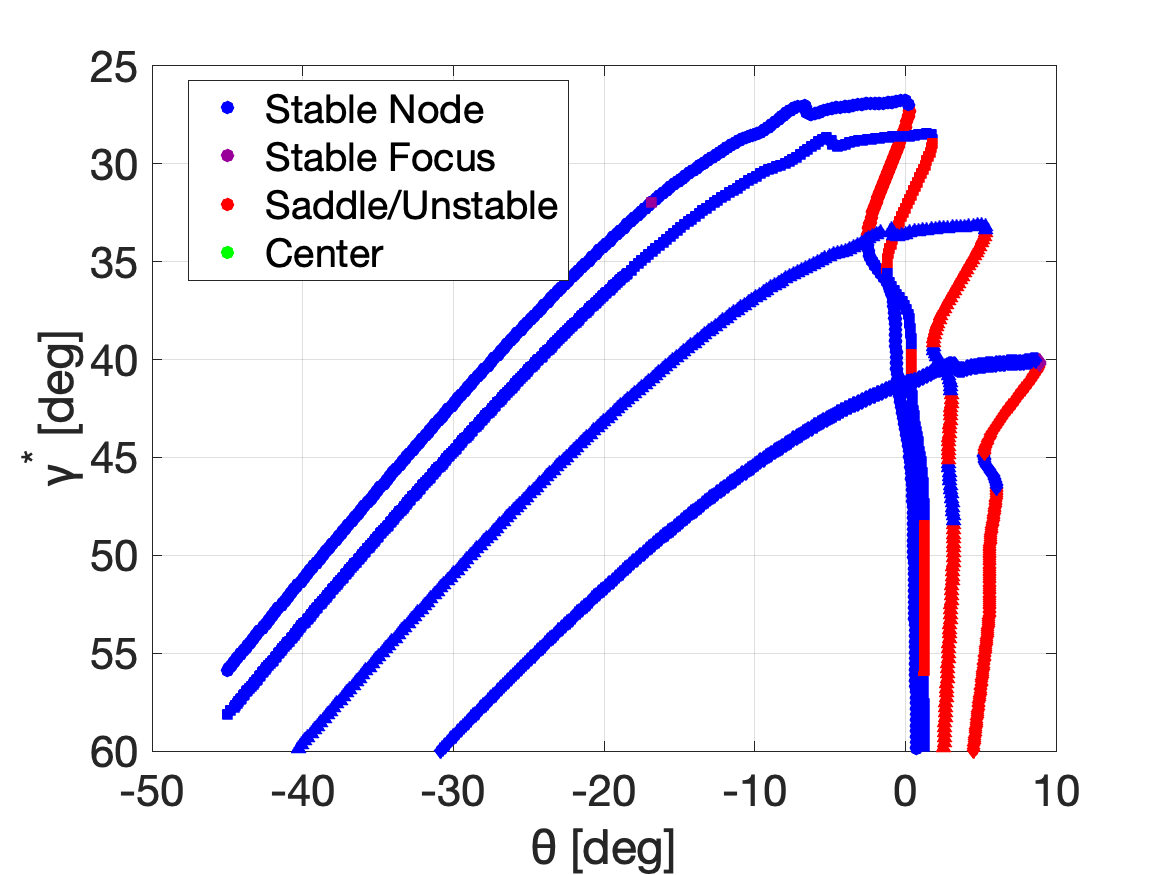}
    \caption{NACA model}
    \label{fig:NACA_filtered_overlay}
  \end{subfigure}

  \vspace{0em}

  \begin{subfigure}[t]{\textwidth}
    \centering
    \includegraphics[width=0.35\textwidth]{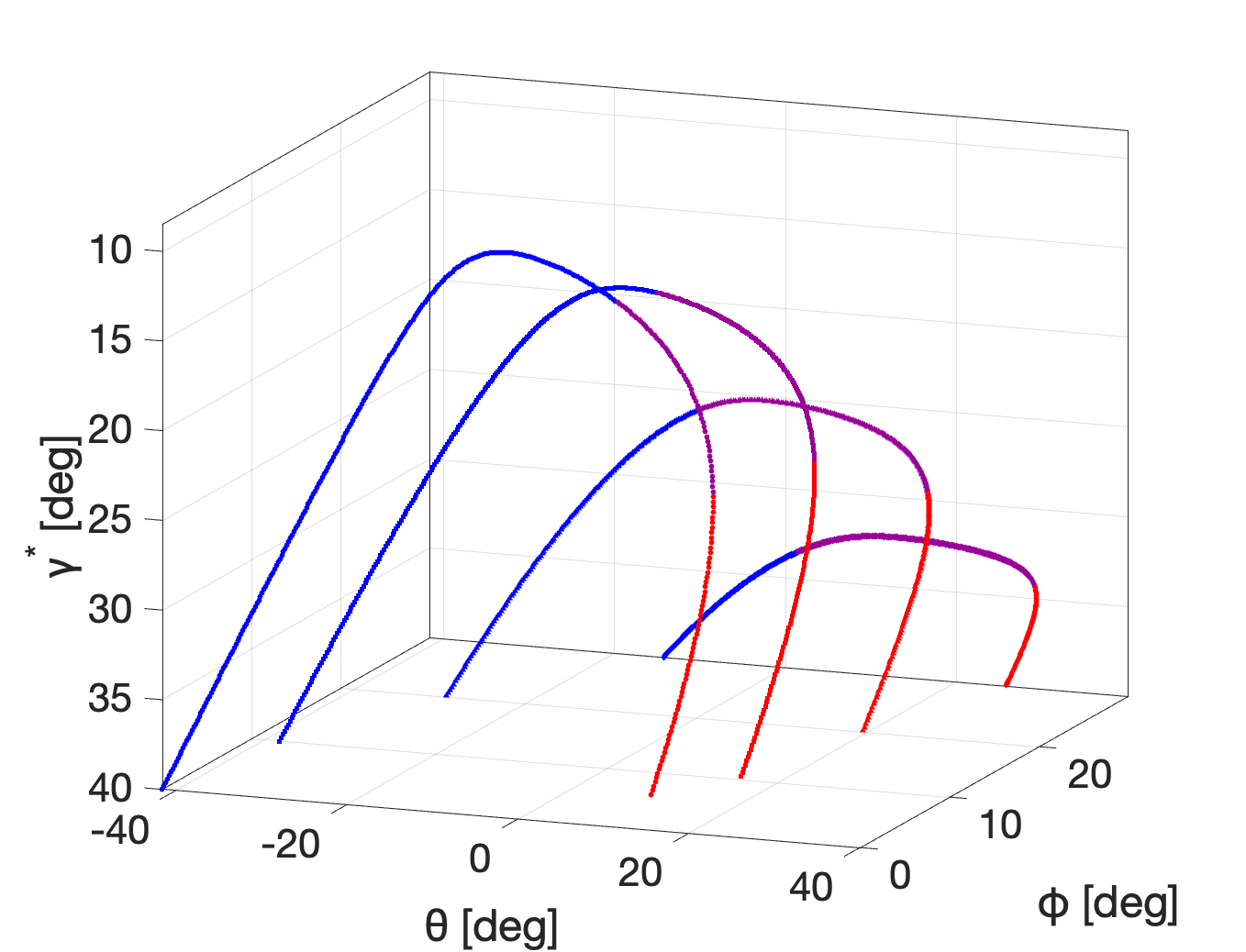}\hfill
    \includegraphics[width=0.35\textwidth]{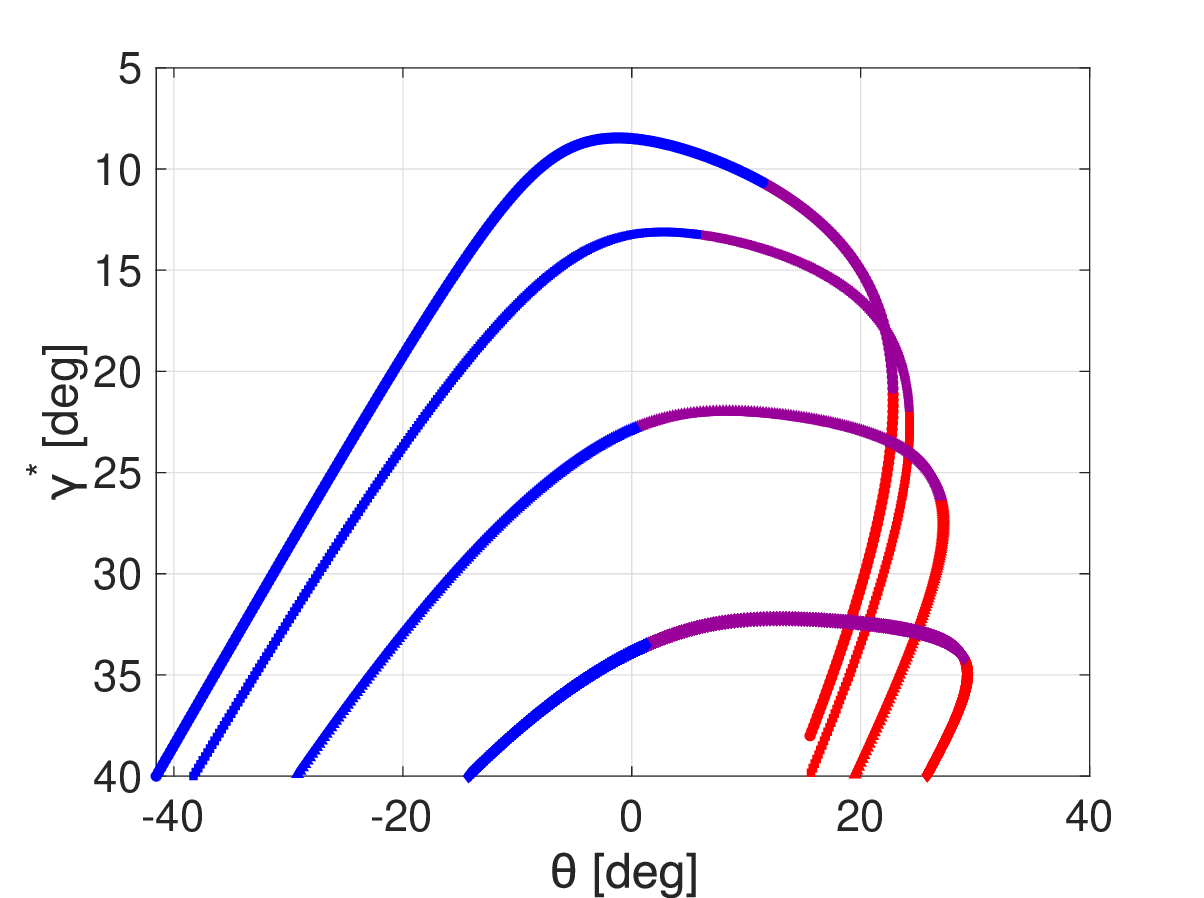}
    \caption{Zimmerman model}
    \label{fig:Zimm_filtered_overlay}
  \end{subfigure}

  \caption{Bifurcation diagrams for the three airfoils. 
  Equilibrium branches are colored by stability: blue for stable nodes, purple for stable foci, and red for unstable (saddle-type) equilibria.
  Yaw is fixed at $\psi=0$, and each curve shows a slice of the bifurcation surface in $(\gamma^*,\theta)$ at a fixed roll angle. Roll $\phi$ varies from $0^{\circ}$ to
  $30^{\circ}$ in increments of $10^{\circ}$,
  revealing how the number, position, and stability of equilibrium glide states depend on body orientation. }
  \label{fig:filtered_overlays_combined}
\end{figure}

Having established the structure of a single, smoothly varying equilibrium surface for the Zimmerman airfoil, we now compare the global bifurcation diagrams of the snake, NACA, and Zimmerman airfoils. 
These diagrams are generated by sweeping $(\theta,\phi)$, keeping $\psi=0$, while solving \eqref{equilibrium} for $(\gamma^*,\sigma^*)$ and classifying stability from the Jacobian \eqref{jacobian_cart}.
For comparison with prior work in \cite{yeaton2017global,nave2019global}, in Figure~\ref{fig:filtered_overlays_combined} we examine sectional plots of $\gamma^{*}$ versus $\theta$ at fixed values of roll, taking $\phi$ from $0^{\circ}$ to $30^{\circ}$ in $10^{\circ}$ increments. 
These slices illustrate the number, organization, and stability transitions of the equilibrium branches across the $(\theta,\phi)$ parameter plane.

\paragraph{Snake Airfoil.}
The snake exhibits clear multistability in Figure \ref{fig:filtered_overlays_combined}(a),  
with overlapping equilibrium branches in $(\theta,\gamma^*)$ projections. 
This means that for a given pitch parameter $\theta$, there could be multiple equilibria. 
On the scale of the given figure, this is only seen for $\theta > -5^{\circ}$.
For each fixed $\phi$, there's a bifurcation point in $\theta$, indicating a saddle-node-type transition.  
Increasing~$\phi$ shifts the entire family of equilibria and produces steeper stable glide angles (i.e., increasing $\gamma^*$),  and moves the bifurcation point to larger $\theta$.

\paragraph{NACA Airfoil.}
Among the three airfoils, the NACA 0012 exhibits the most pronounced multistability. 
As shown in Figure~\ref{fig:filtered_overlays_combined}(b), at up to six distinct equilibrium branches coexist for certain ranges of pitch angle~$\theta$. 
These branches alternate in stability---stable, then unstable, then stable again---producing a layered sequence of saddle-node folds and rapidly shifting basins of attraction. 
Unlike the snake,  the NACA's structure represents genuine multiplicity of equilibria in orientation space. 
Roll variations shift these branches, and create new bifurcations changing multi-stable structure, indicating that the NACA airfoil has complex equilibrium organization.

\paragraph{Zimmerman Airfoil.}
The Zimmerman airfoil exhibits a bifurcation structure broadly similar to the snake, with several equilibrium branches visible in the $(\theta,\gamma^*)$ projection and alternating regions of stability. 
However, the stable branches occur at considerably smaller glide angles, indicating much shallower equilibrium descent. 
The folds are smoother and slightly less sharply curved than in the snake, and the projection overlap is more modest.
Although multistability is present, the Zimmerman airfoil has a broad region of stable, shallow glide states, reflecting its high $C_L/C_D$ ratio and gentle stall behavior.

\paragraph{Stability Structure in the Full Pitch-Roll Parameter Space.}
Across the three airfoils, the bifurcation diagrams show that equilibrium stability is not governed solely by pitch $\theta$ 
but can also change substantially as roll $\phi$ varies.
For the snake and Zimmerman airfoils, a single stable branch at low $\phi$ may transition to a focus or lose stability entirely as roll increases, revealing that roll-induced bifurcations are as important as pitch-dependent folds.
The NACA 0012 displays the strongest form of this behavior: its multiple coexisting equilibria undergo repeated stability switches in both $\theta$ and $\phi$, producing alternating node-saddle structure across the orientation parameter space.
The diagrams make clear that robustness is governed by the stability structure in the full $(\theta,\phi)$ space, not by pitch alone. 
The three airfoils differ  in how their equilibria gain or lose stability under changes in roll.

\section{Terminal Velocity Manifold and Separatrix Structure}

Before analyzing gliding efficiency, it is essential to introduce the global invariant structures that organize the velocity dynamics.
Across all airfoils, the flow in velocity space is shaped by two key geometric objects: the terminal velocity manifold, an attracting slow-fast surface, and a separatrix, the invariant surface that partitions shallow, lift-dominated glides from steep, drag-dominated descent.
Together they form the phase-space backbone of non-equilibrium gliding.

\subsection{Terminal Velocity Manifold}

In classical fluid mechanics, the \emph{terminal velocity} is the constant vertical speed attained when gravity and aerodynamic drag balance.  
In the present three-dimensional glider model, this concept generalizes to a two-dimensional \emph{terminal velocity manifold} (TVM): an attracting invariant surface in velocity space $(v_1,v_2,v_3)$.

The TVM acts as a global organizing structure.  
Trajectories collapse rapidly toward it in the direction normal to the surface and subsequently evolve slowly along it toward an equilibrium glide state.  
This separation of timescales---fast transient convergence followed by slow drift tangential to the surface---is characteristic of a normally hyperbolic invariant manifold.  
The TVM therefore forms a global attractor in phase space: trajectories are drawn toward it and evolve along it, but
do not cross it on their way to an ultimate stable equilibrium point.

Although the TVM lies close to the $v_3$-nullcline (the surface where $\dot v_3=0$), the two do not coincide.  
We compute the TVM directly using the bisection method of \cite{nave2019global}, which identifies the attracting surface without relying on nullcline approximations.

Figure~\ref{fig:TVM_combined} illustrates the TVM  for the snake, NACA, and Zimmerman airfoils.  
Each row shows (left) the full two-dimensional TVM in 3D velocity space  (blue surface), (top right) its projection onto the $(v_1,v_2)$ plane, and (bottom right) the temporal evolution of $(v_1,v_2,v_3)$ for a typical initial condition, highlighting the fast-slow structure.  
Across all three airfoils, trajectories rapidly collapse onto the TVM before drifting along it toward a terminal equilibrium glide state.

The TVM reflects distinct descent characteristics across the three airfoils.
For the snake profile, the TVM corresponds to relatively low descent speeds (i.e., $|v_3|$ is not large), whereas both the NACA and Zimmerman airfoils exhibit higher descent speeds along their TVMs.
For the nominal orientation $(\theta,\phi,\psi)=(0^\circ,0^\circ,0^\circ)$, each airfoil possesses more than one equilibrium in the $(v_1,v_3)$ plane: a shallow, stable glide with the largest horizontal velocity component $v_1$, and a steeper unstable equilibrium with reduced $v_1$.
Trajectories that fail to reach the shallow stable glide may instead settle into a steep, less efficient glide and in some cases even converge to an equilibrium with negative $v_1$.
This raises a central question: what dynamical structure divides the desirable (shallow) glide states from the steep, drag-dominated ones?

\begin{figure}[H]
  \centering
  \vspace*{-1cm}
   \hspace*{0cm}
  \begin{subfigure}{\textwidth}
    \centering
    \begin{minipage}[t]{0.5\linewidth}
      \includegraphics[width=\linewidth]{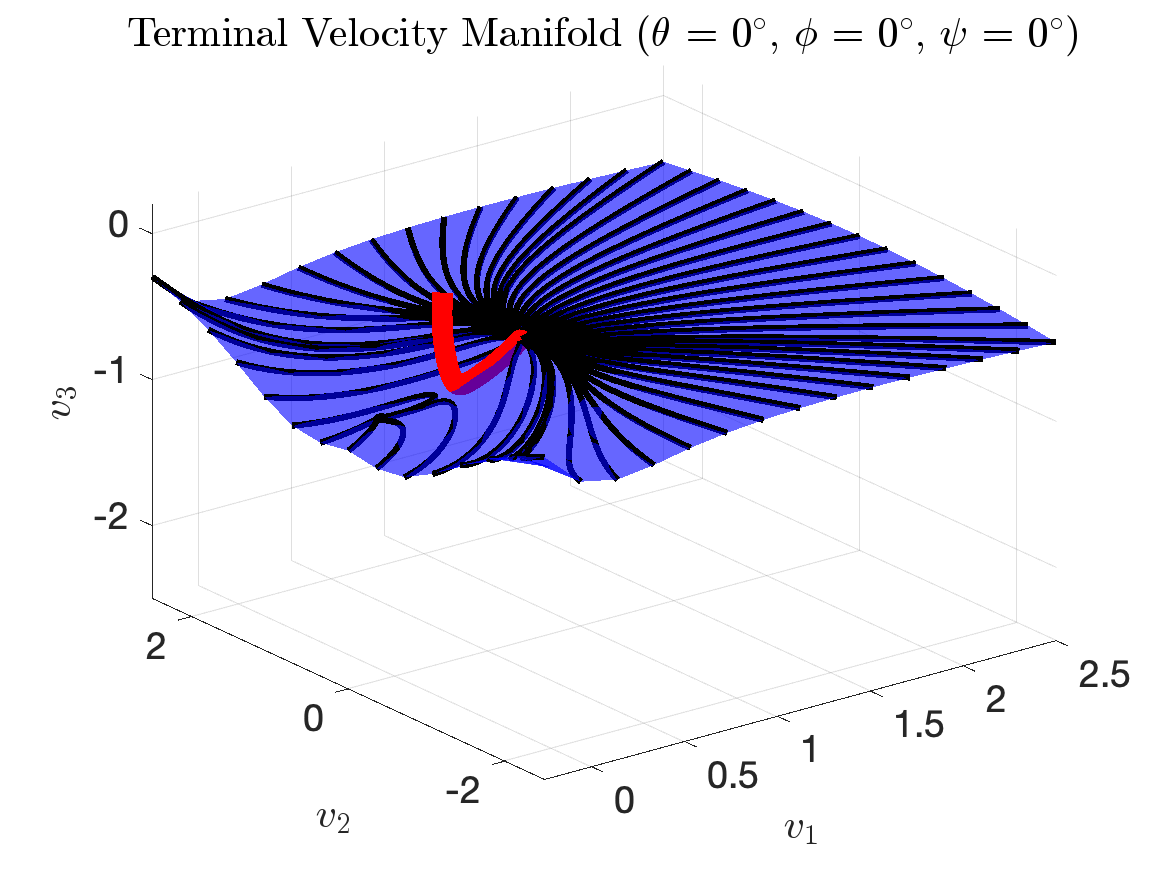} 
    \end{minipage}%
    \hfill
    \raisebox{3cm}{%
    \begin{minipage}[t]{0.27\linewidth}
      \includegraphics[width=\linewidth]{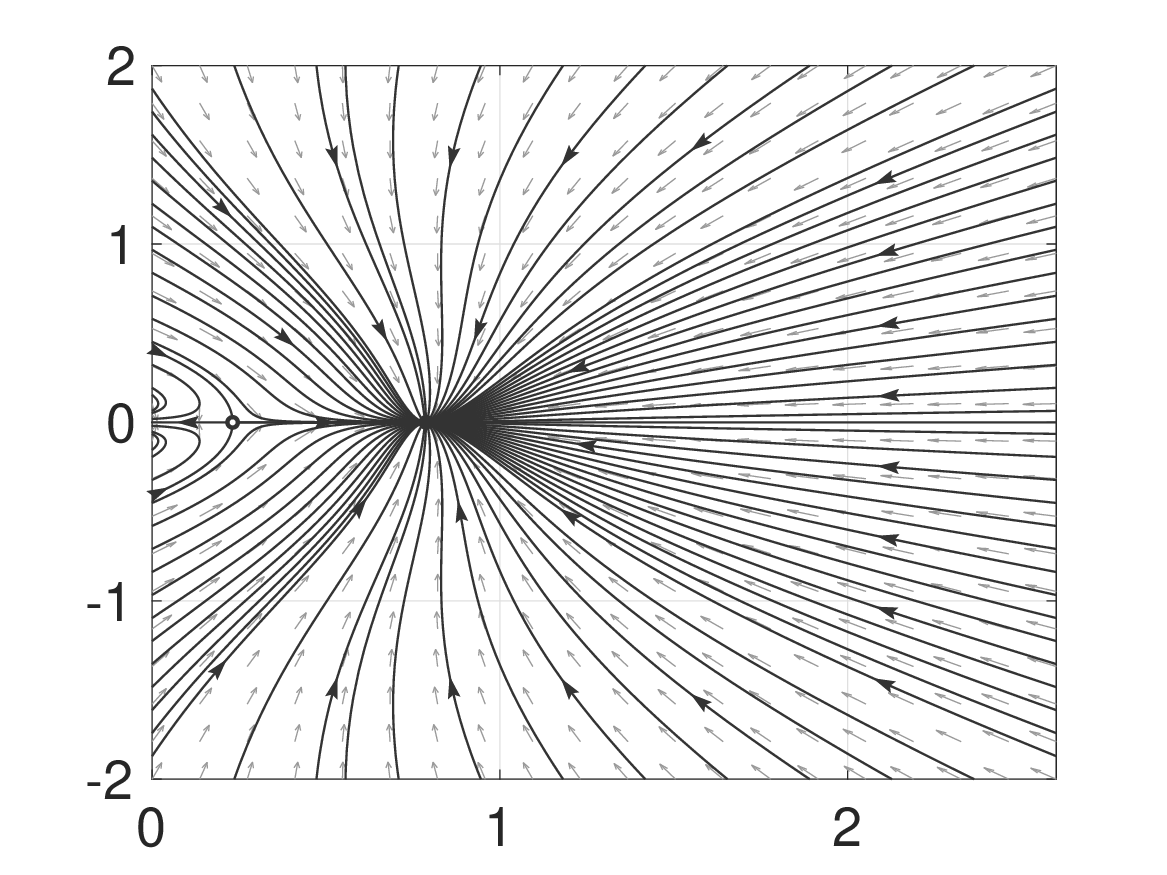}  
      \includegraphics[width=\linewidth]{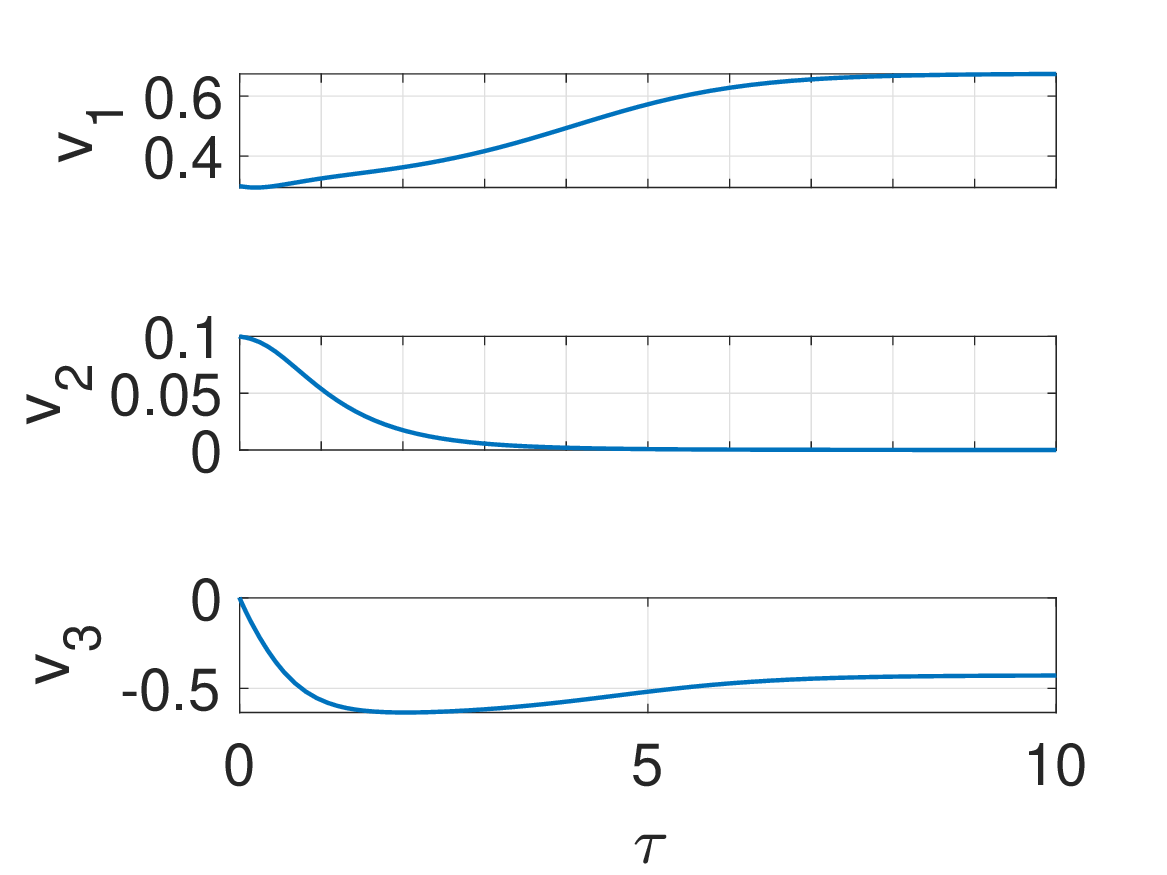} 
    \end{minipage} }
    \caption{}
    \label{fig:tvm_3D}
  \end{subfigure}


  \hspace*{0cm}
  \begin{subfigure}{\textwidth}
    \centering
    \begin{minipage}[t]{0.5\linewidth}
      \includegraphics[width=\linewidth]{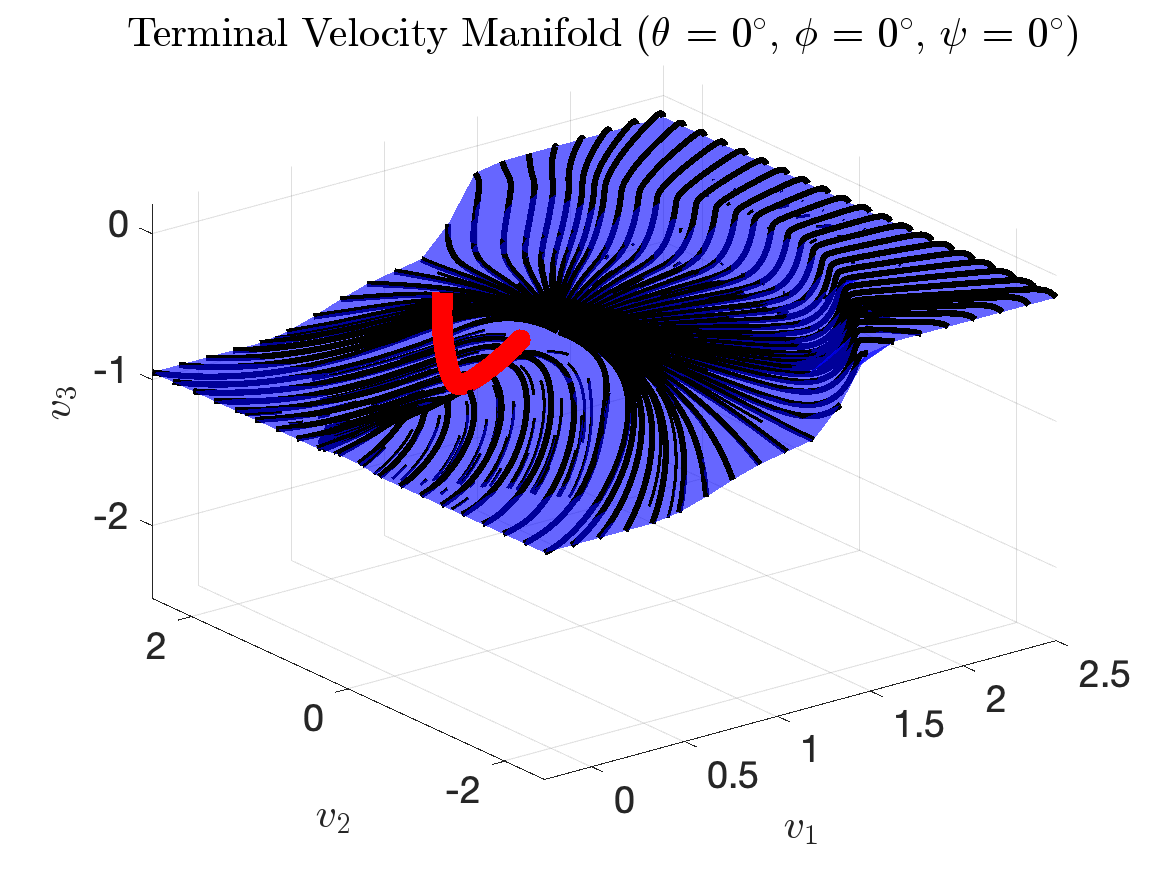}
    \end{minipage}%
    \hfill
    \raisebox{3cm}{%
    \begin{minipage}[t]{0.27\linewidth}
      \includegraphics[width=\linewidth]{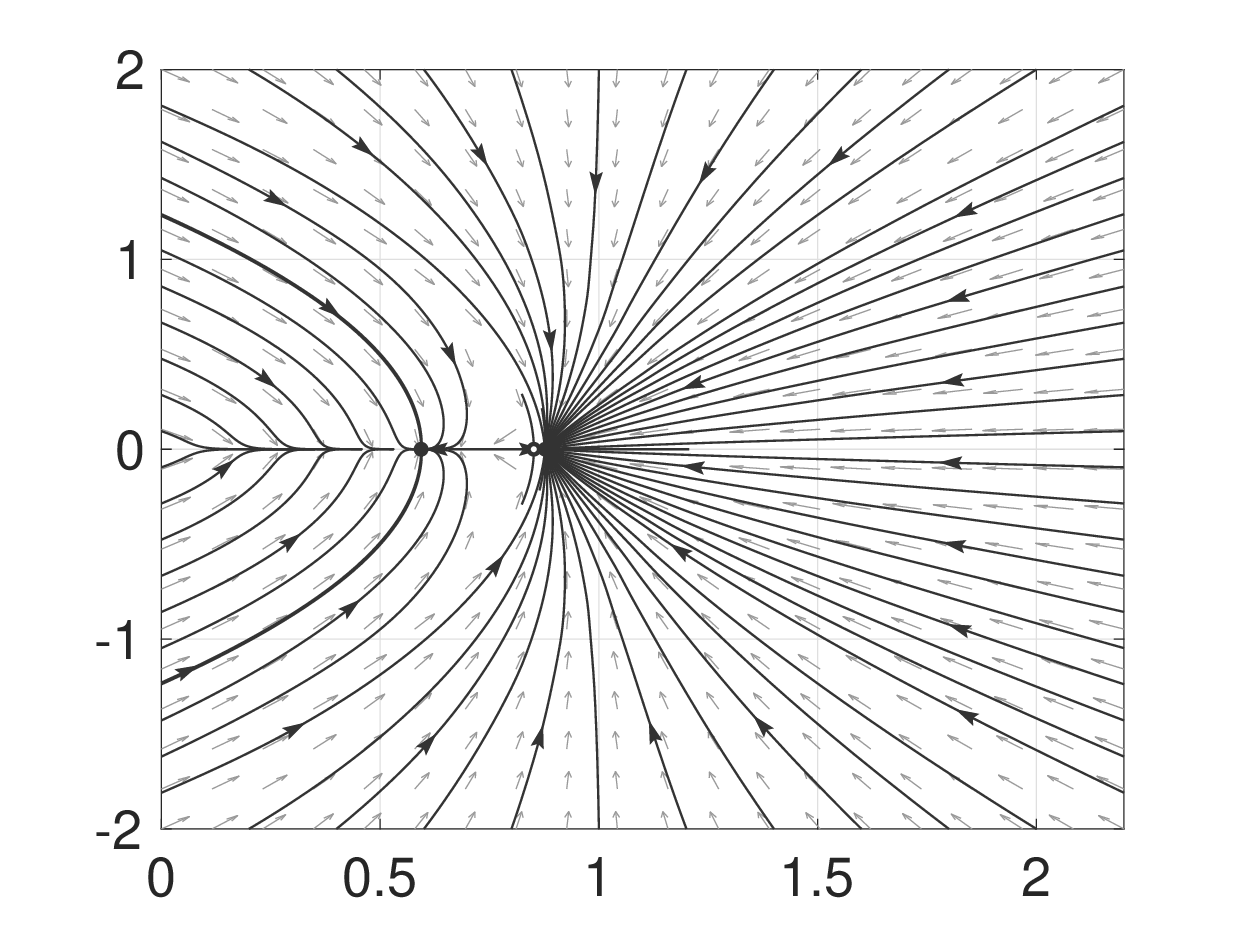}
      \includegraphics[width=\linewidth]{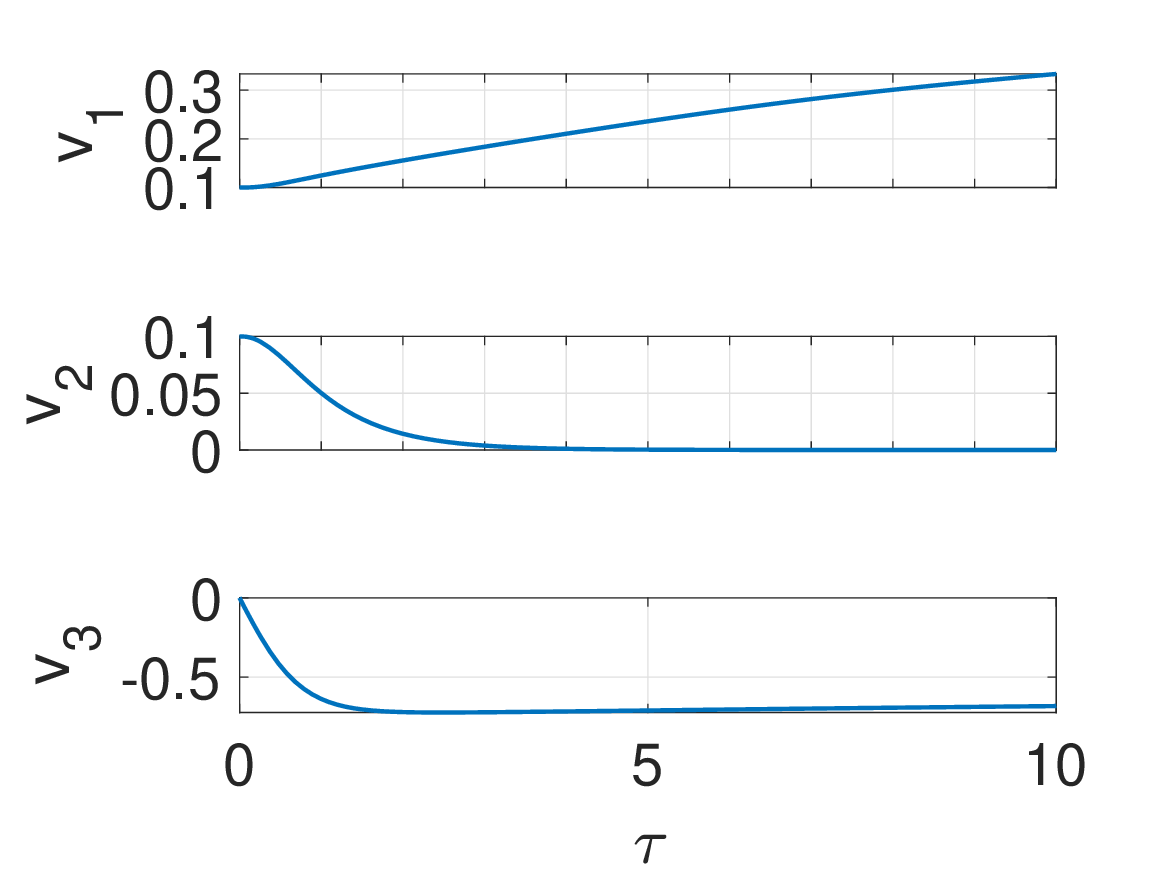} 
    \end{minipage} }
    \caption{}
    \label{fig:phase_port}
  \end{subfigure}


   \hspace*{0cm}
  \begin{subfigure}{\textwidth}
    \centering
    \begin{minipage}[t]{0.5\linewidth}
      \includegraphics[width=\linewidth]{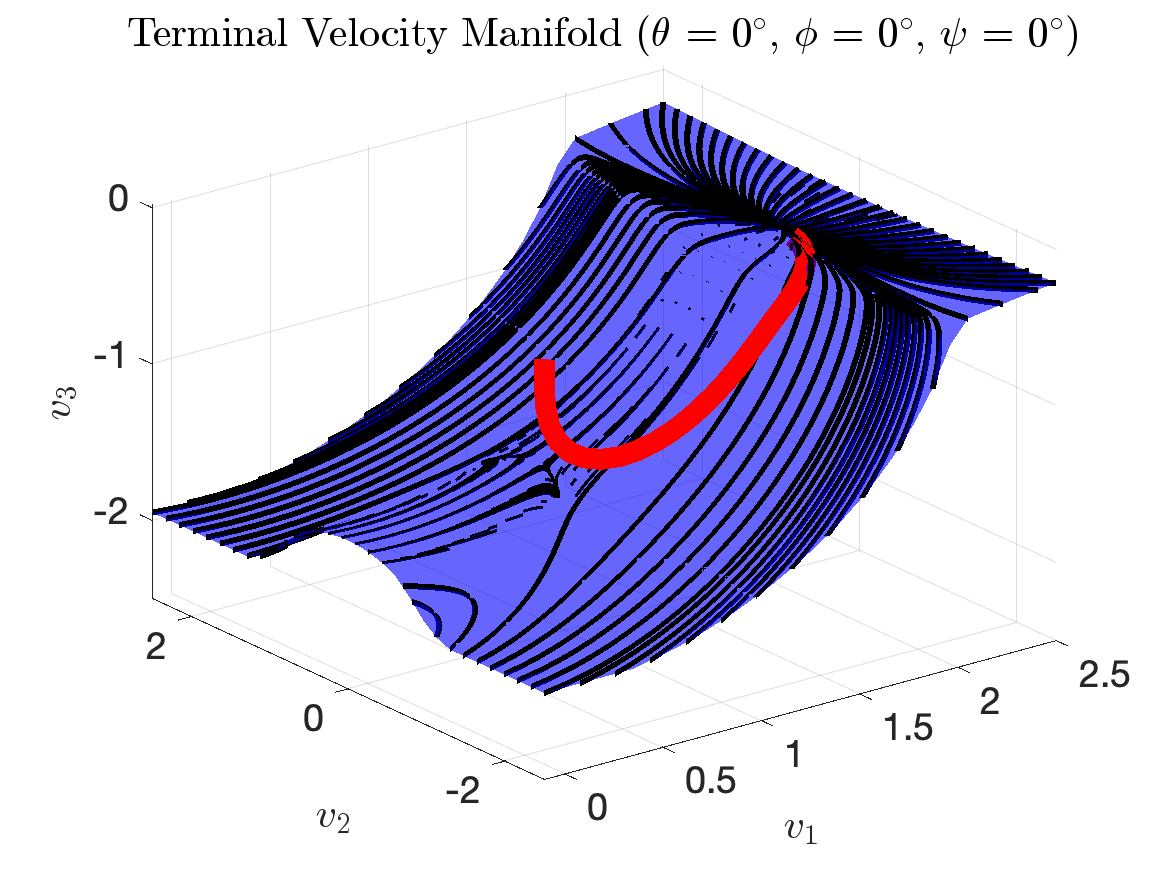}
    \end{minipage}%
    \hfill
    \raisebox{3cm}{%
    \begin{minipage}[t]{0.27\linewidth}
      \includegraphics[width=\linewidth]{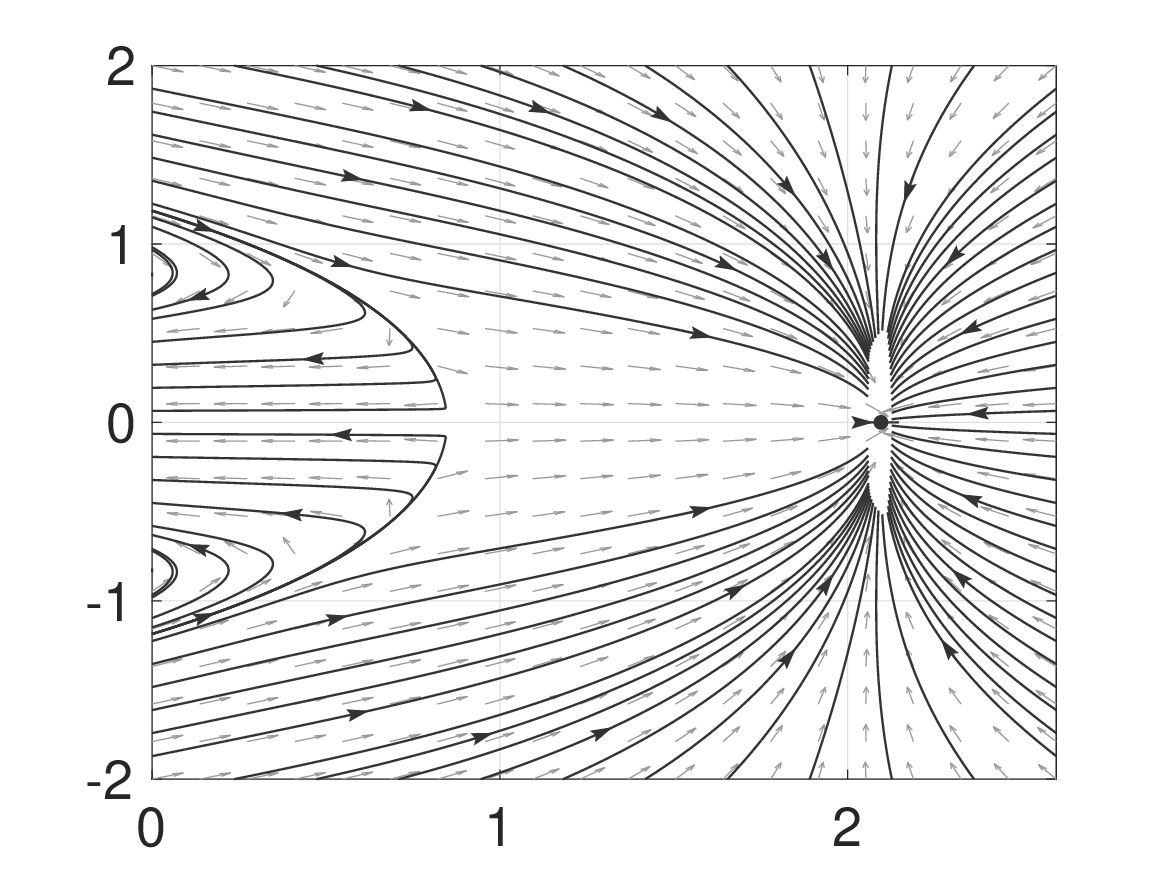}
      \includegraphics[width=\linewidth]{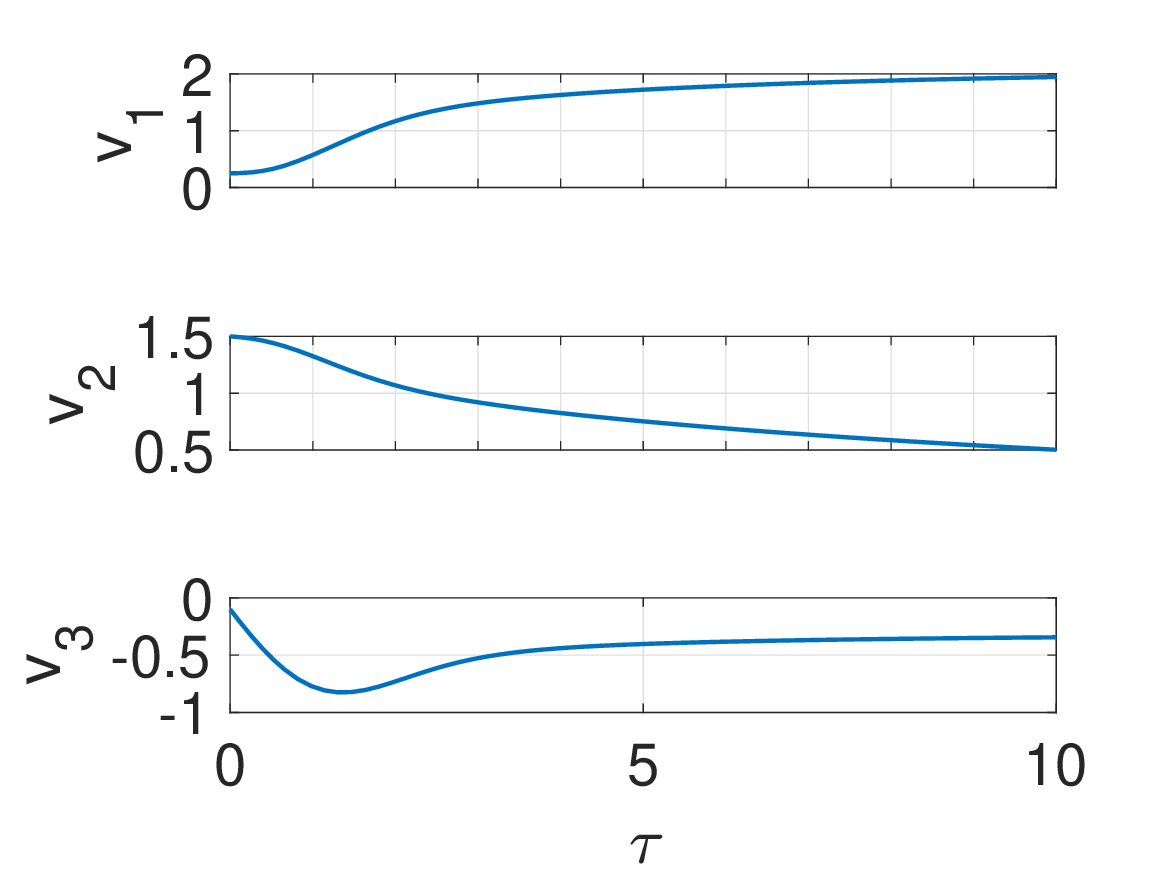} 
    \end{minipage} }
    \caption{}
    \label{fig:temporal_NS}
  \end{subfigure}
 \vspace*{-0.8cm}
  \caption{From top to bottom in order : (a) Snake airfoil, (b) NACA 0012 airfoil, and (c) Zimmerman airfoil. Phase portraits projected on the TVM are in the top right corners (using \cite{Zhang2026PhasePortraitPlotter}).}
  \label{fig:TVM_combined}
\end{figure}

\subsection{Separatrix Between Shallow Glides and Steep Descent}

In planar systems, a separatrix is an invariant trajectory through a saddle equilibrium that divides qualitatively different dynamical regimes;
for example,  the heteroclinic orbit separating oscillatory from rotational motion in the simple pendulum \cite{Strogatz2018,Perko2001}. 
In higher dimensions, this generalizes to a codimension-one invariant \emph{separatrix surface}, typically the stable or unstable manifold of a saddle-type equilibrium.  
Such  surfaces partition the three-dimensional phase space into distinct basins of attraction \cite{Wiggins1992}.

In the present 3D glider model, the relevant separatrix is the two-dimensional stable manifold of the saddle-like unstable equilibrium located on the TVM. 
This surface intersects the TVM and partitions the velocity space into two dynamical regimes:
one leading to shallow, lift-dominated glide states and the other to steep, drag-dominated descent.
Because the separatrix is invariant, trajectories cannot cross it for a fixed orientation.
However, as the body orientation $(\theta,\phi,\psi)$ changes, the separatrix itself can shift, altering which initial conditions fall into each regime.

The three airfoils show distinct separatrix geometries and relationships to the TVM and equilibria locations; see 
Figure~\ref{fig:TVM_combined}.

\begin{figure}[H]
  \centering
  \vspace*{-1cm}
   \hspace*{0cm}
  \begin{subfigure}{\textwidth}
    \centering
    \begin{minipage}[t]{0.5\linewidth}
      \includegraphics[width=\linewidth]{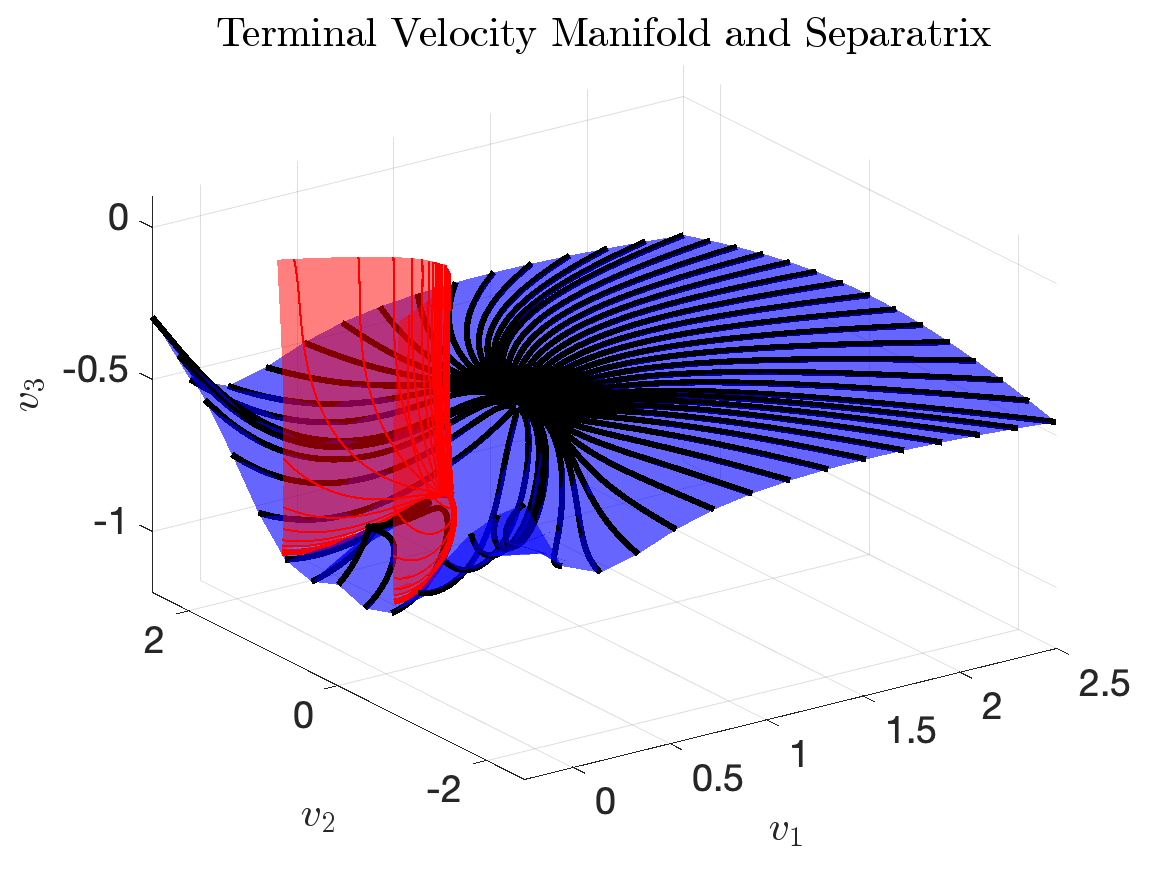} 
    \end{minipage}%
    \hfill
    \begin{minipage}[t]{0.5\linewidth}
      \includegraphics[width=\linewidth]{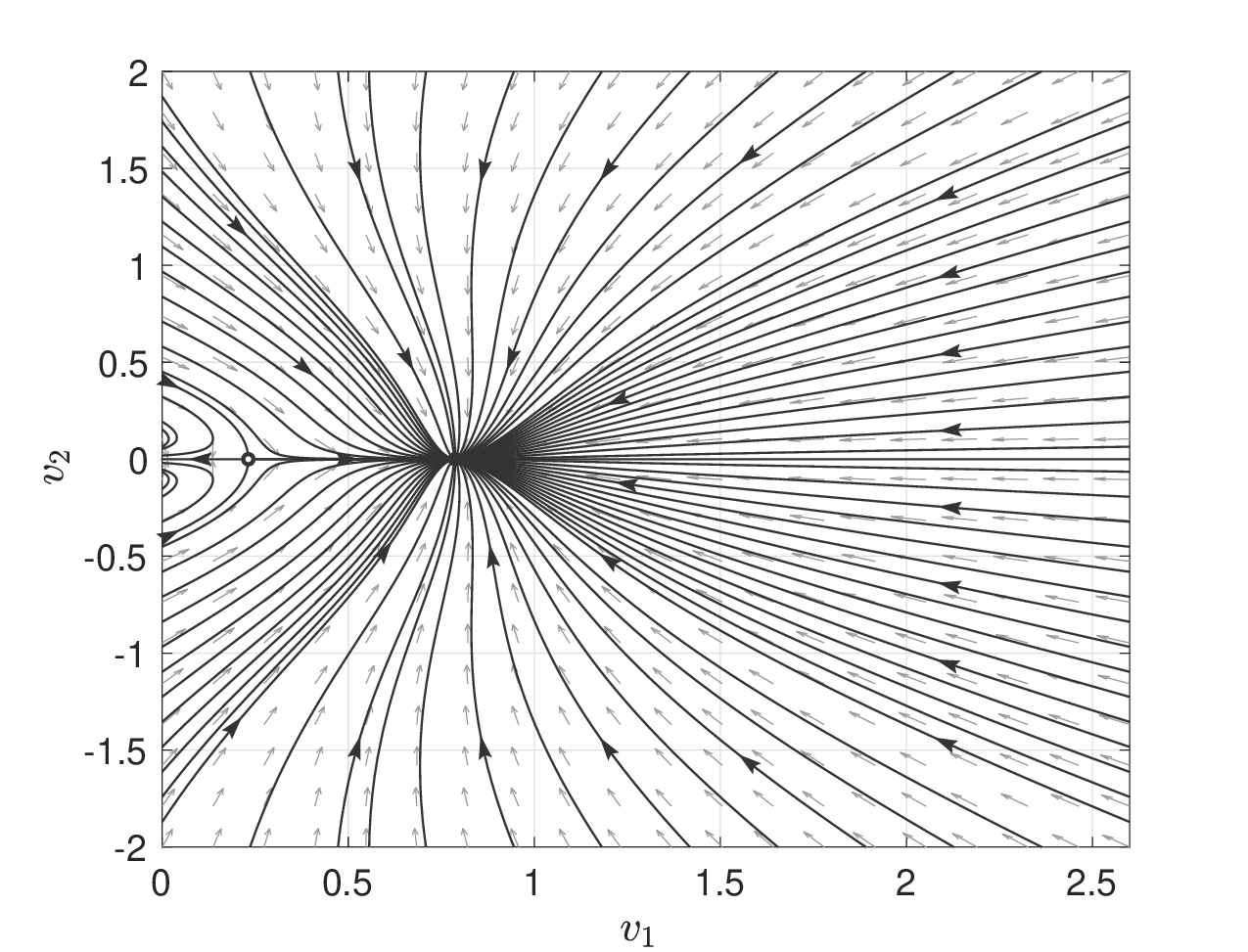}   
    \end{minipage} 
    \label{fig:tvm_3D}
  \end{subfigure}


  \hspace*{0cm}
  \begin{subfigure}{\textwidth}
    \centering
    \begin{minipage}[t]{0.5\linewidth}
      \includegraphics[width=\linewidth]{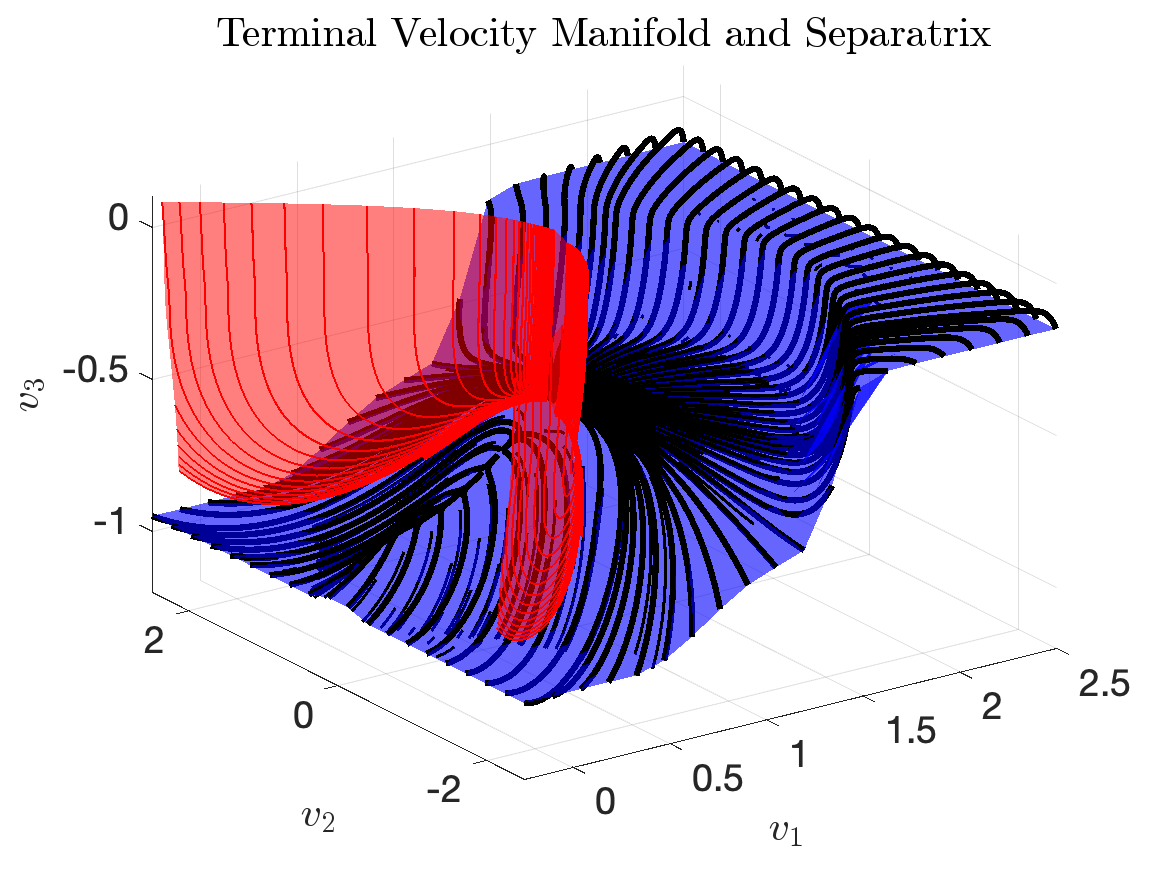}
    \end{minipage}%
    \hfill
    \begin{minipage}[t]{0.5\linewidth}
      \includegraphics[width=\linewidth]{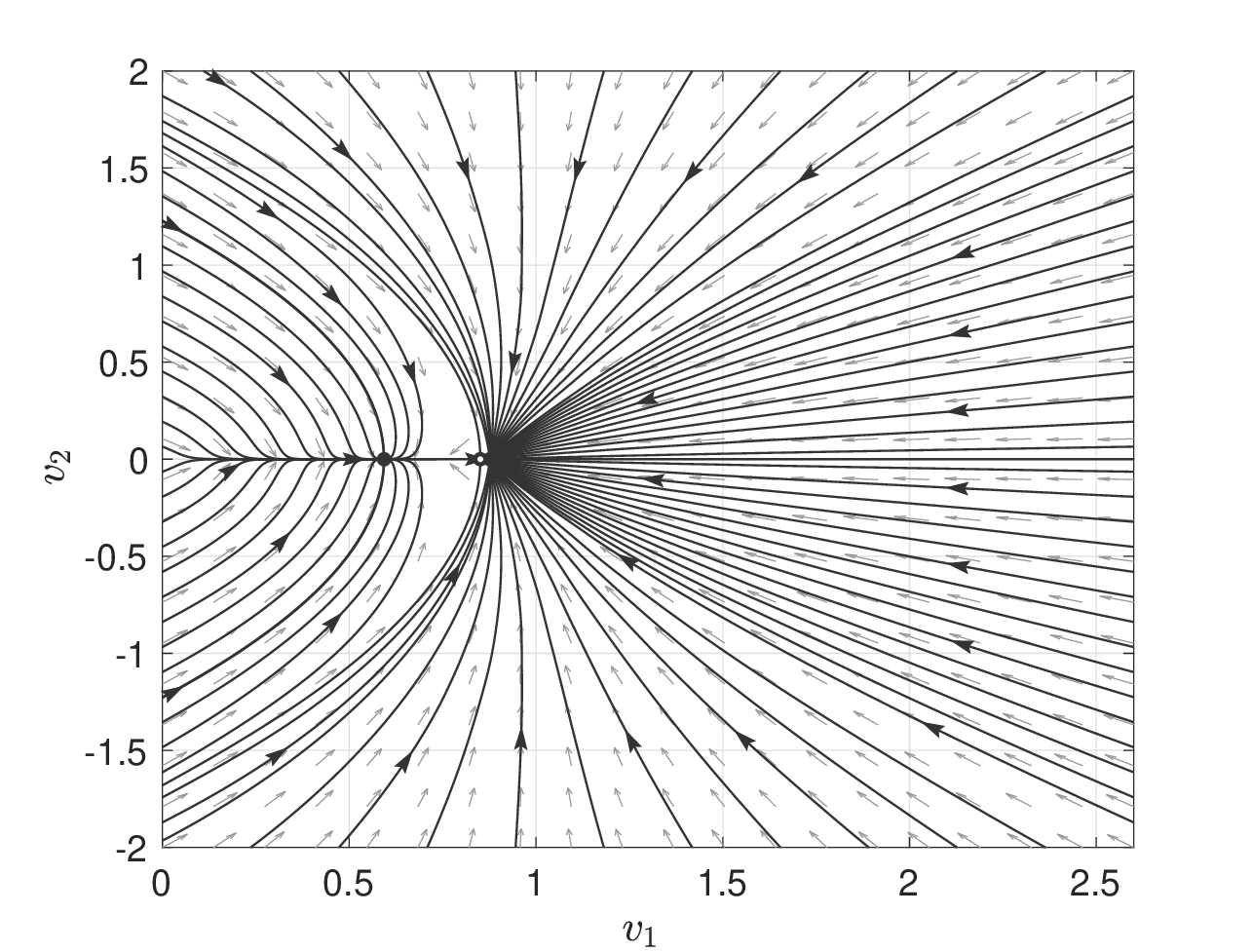}
    \end{minipage} 
    \label{fig:phase_port}
  \end{subfigure}

    \hspace*{0cm}
  \begin{subfigure}{\textwidth}
    \centering
    \begin{minipage}[t]{0.5\linewidth}
      \includegraphics[width=\linewidth]{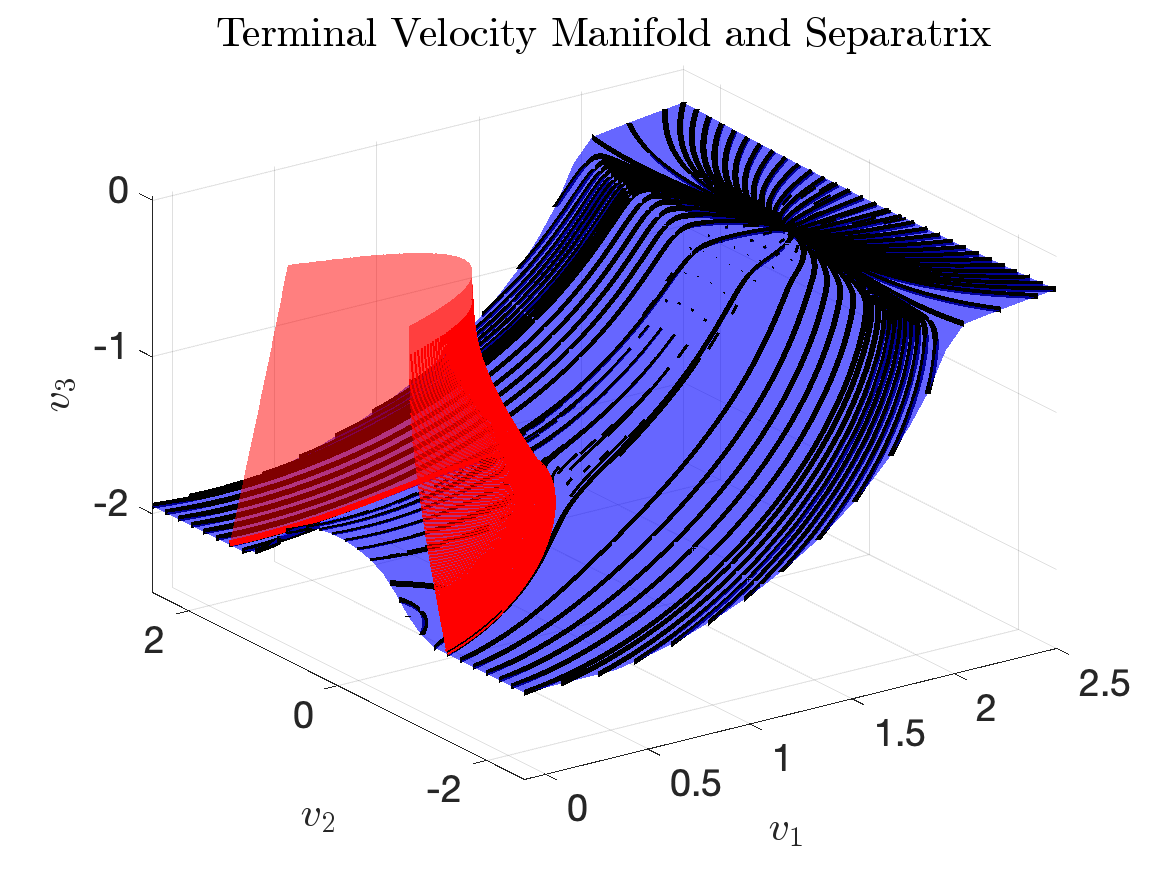}
    \end{minipage}%
    \hfill
    \begin{minipage}[t]{0.5\linewidth}
      \includegraphics[width=\linewidth]{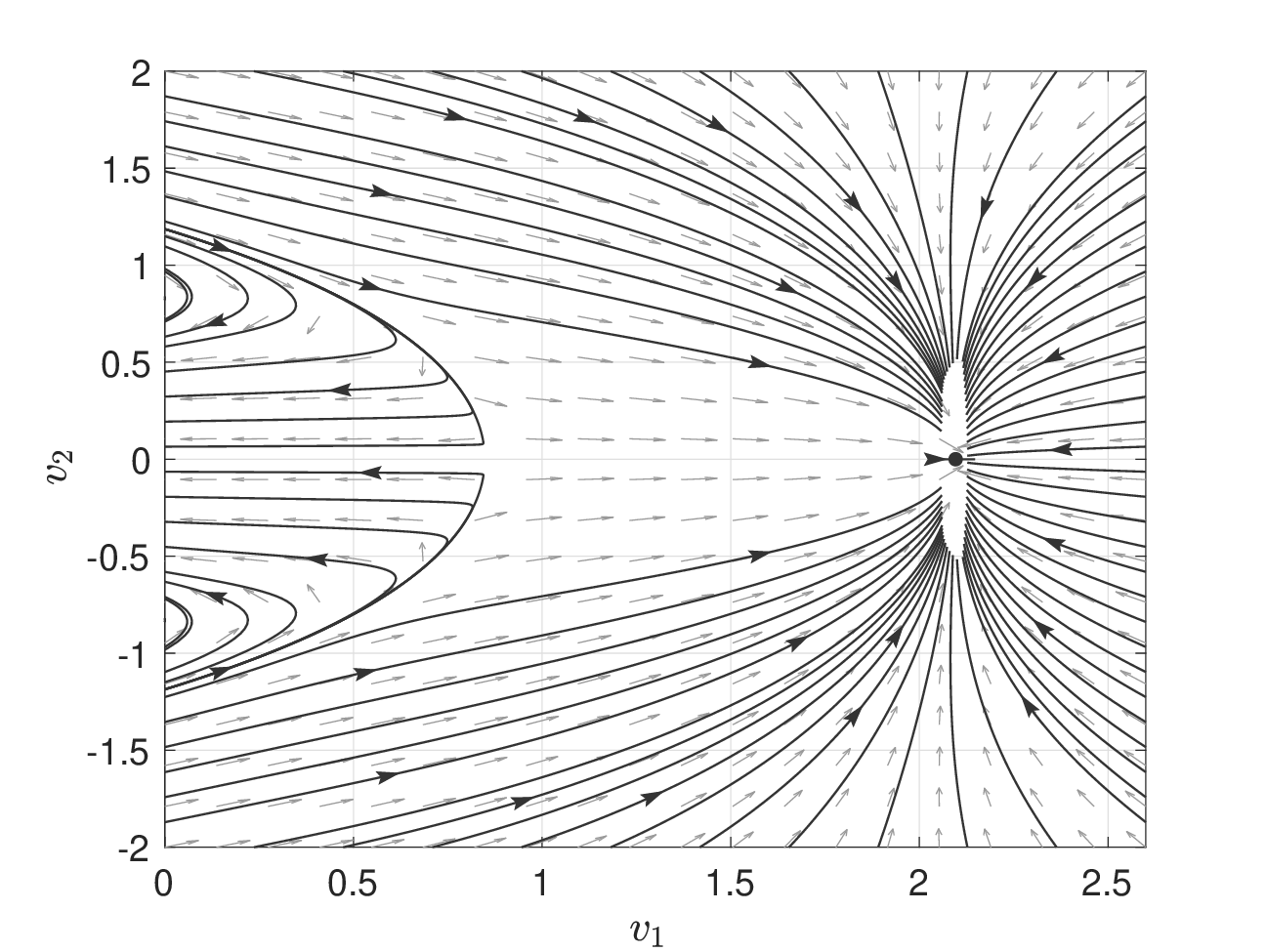}
    \end{minipage} 
    \label{fig:phase_port}
  \end{subfigure}


 \vspace*{-0.5cm}
  \caption{TVM for the three airfoils from top to bottom:  snake model, NACA 0012, Zimmerman.}
  \label{fig:TVM_combined}
\end{figure}

\begin{figure}[H]
  \centering
  \begin{subfigure}{\textwidth}
    \centering
    \begin{minipage}[t]{0.7\linewidth}
      \includegraphics[width=\linewidth]{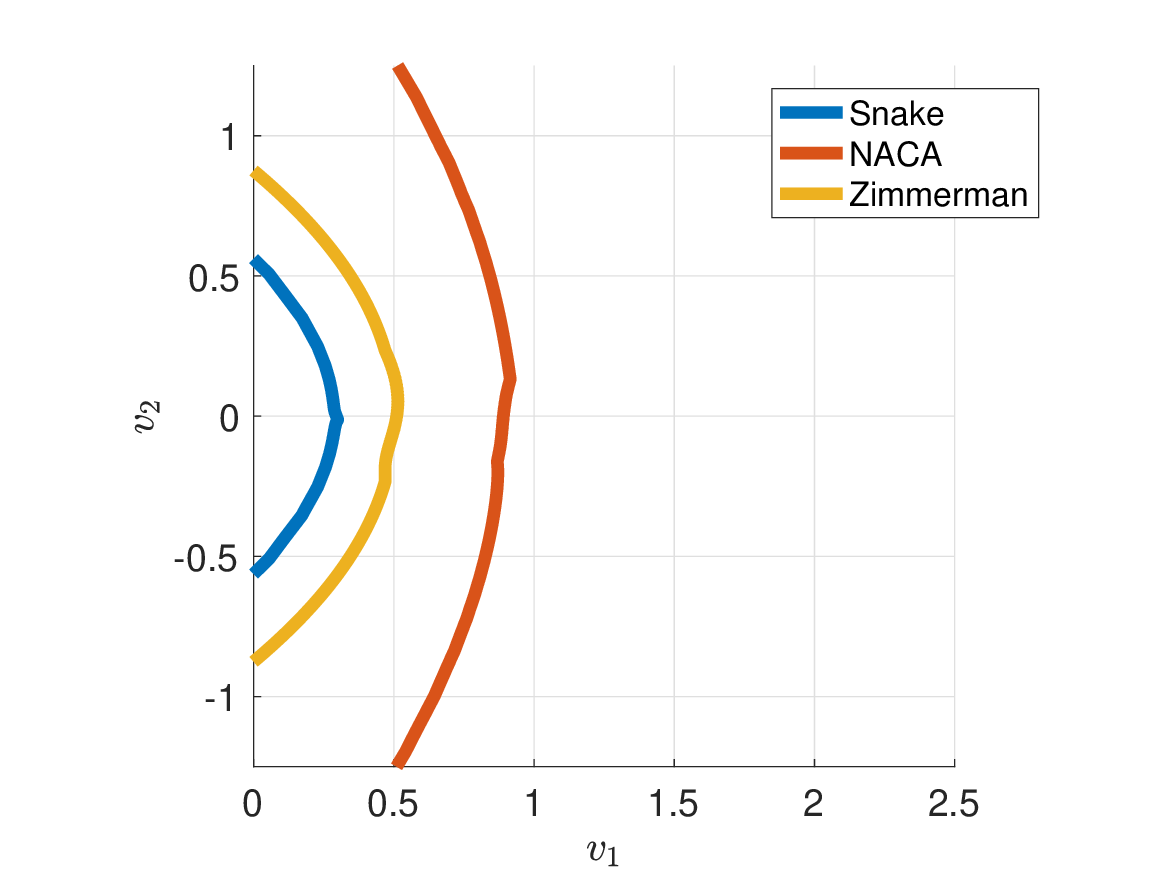} 
    \end{minipage}%
    \hfill
    \begin{minipage}[t]{0.7\linewidth}
      \includegraphics[width=\linewidth]{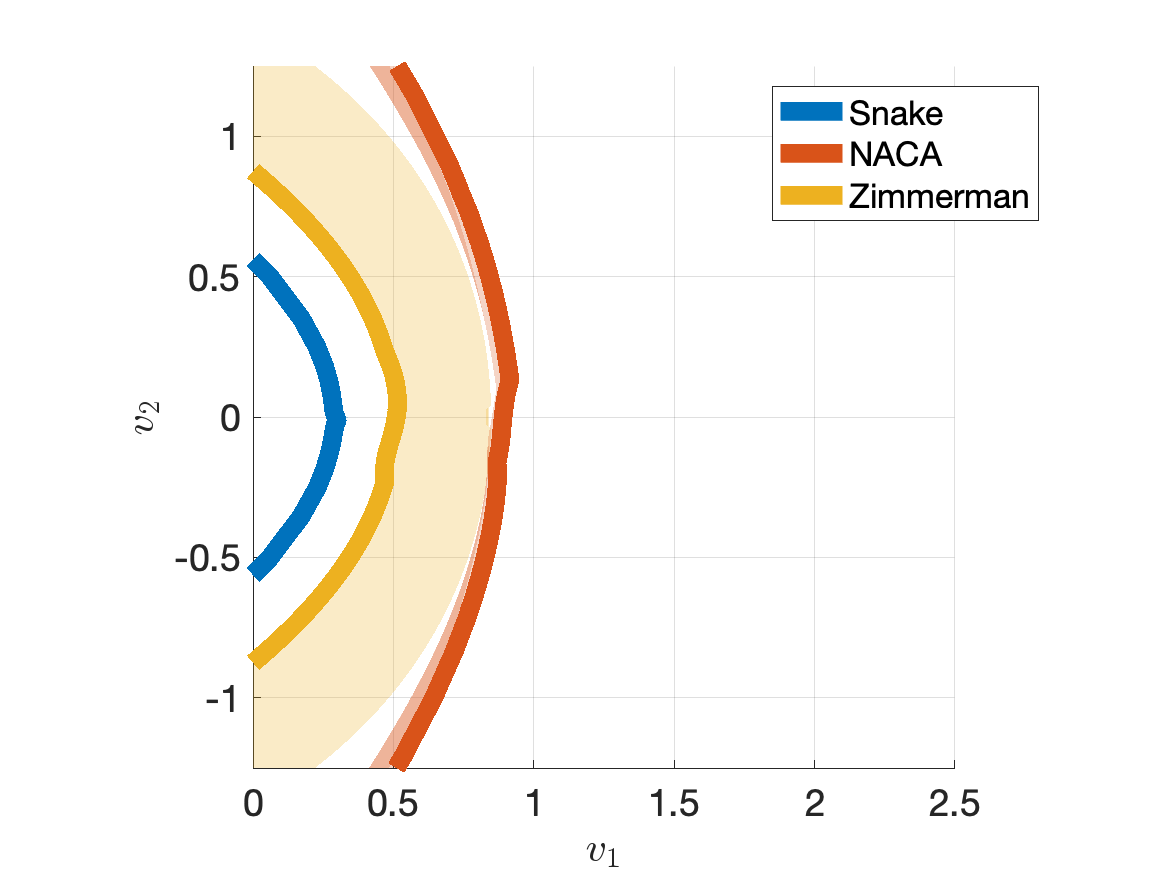}   
    \end{minipage} 
  \end{subfigure}

  \caption{Separatrix for the three airfoils in the $v_1$ - $v_2$ plane ( top ) and in the full 3D plane ( bottom) . The top figure shows the intersection of the separatrix with $v_3 = 0$, and the bottom figure shows the full 3D separatrix curves for all three planforms.  }
  \label{fig:sep_top}
  
  \end{figure}

\paragraph{Snake Airfoil.}
The snake's separatrix in Figure~\ref{fig:TVM_combined}(a) has several generic features, forming a broadly parabolic sheet (red surface) in $(v_1,v_2,v_3)$-space.
We plot only the portion lying above the TVM, since this is the dynamically relevant region for gliding animals that begin with a horizontal leap, e.g., a gliding animal jumping from a tree.
The key diagnostic, therefore, is the intersection of the separatrix with the plane $v_3 = 0$, as seen in Figure \ref{fig:sep_top}(a) which represents all purely horizontal initial jump states.
Initial conditions lying outside this separatrix footprint converge to the shallow, desirable equilibrium glide; initial conditions inside it instead evolve toward a steeper, drag-dominated descent glide state.

For a fixed horizontal launch speed $v_h=\sqrt{v_1^2 + v_2^2}$, the separatrix reveals that the most efficient trajectory is obtained when the jump is directed primarily along the forward axis ($v_1$).
Departures in the lateral direction ($v_2$) place the initial state closer to, or even inside, the separatrix, increasing the likelihood of falling into the steep-descent regime.
Thus, the snake'd separatrix geometry provides a clear predictor of which horizontal launch directions lead to efficient gliding and which require active body modulation to avoid rapid loss of altitude.

\paragraph{NACA  Airfoil.}
The NACA airfoil exhibits the widest and most displaced separatrix of the three profiles.
As seen in Figure~\ref{fig:TVM_combined}(b), the separatrix sits very close to the shallow glide on the TVM; they are separated by a small amount in the $v_1$ direction.
This implies that even modest deviations in the initial velocity orientation push trajectories into drag-dominated descent before they can reach the shallow, energetically favorable glide.

The intersection of the separatrix with the horizontal plane $v_3 = 0$ ( Figure \ref{fig:sep_top}(a) ) forms a  large lobe of admissible initial jumps: only carefully aligned forward jumps avoid falling directly into the steep stable glide regime.
Together, these features reflect the complex multistability and sensitivity of the NACA 0012: although capable of efficient glide in principle, its desirable states occupy a small, delicate region of the initial-condition space.

\paragraph{Zimmerman Airfoil.}
The Zimmerman airfoil displays the most favorable TVM-separatrix geometry.
Its separatrix  is compact and positioned far from the shallow-glide equilibrium branch on the TVM.
This shallow glide, characterized by large forward velocity ($v_1 \approx 2$) and relatively small descent speed ($v_3 \approx -0.5$), is the most robust among all three airfoils.

In the horizontal plane $v_3=0$ as seen in Figure \ref{fig:sep_top}(a), the separatrix encloses a relatively narrow parabolic region of ``bad'' initial jumps.
Most forward-oriented velocities lie well outside this basin, ensuring that even imperfect launches rapidly collapse onto the  region of the TVM leading to the shallow glide, rather than drifting into steep descent. As can also be seen in Figure \ref{fig:sep_top}(b), between the three planforms, the Zimmerman separatrix is the one that stood out as time progressed. As we approached  the $v_3 = 0$ plane, the area of the separatrix collapsed.

\paragraph{Implications for Glide Robustness.}
Taken together, the TVM-separatrix geometry provides a dynamical-systems perspective on glide robustness.
Unlike the NACA profile---whose separatrix lies close to its shallow-glide branch, restricting efficient glide to a narrow band of finely tuned initial conditions---the two bio-inspired airfoils (snake and Zimmerman) possess relatively smaller separatrix regions.
This compactness creates a wide buffer around the states leading to shallow equilibrium glide, meaning that a broad range of initial horizontal jump conditions naturally collapse onto the efficient side of the TVM rather than slipping into steep, drag-dominated descent.
Accordingly, both biological airfoils tolerate substantial variation in launch orientation yet still achieve shallow, high-efficiency glides.
In contrast, the NACA profile's relatively large  separatrix region implies reduced robustness and greater sensitivity to initial conditions.
Thus, separatrix geometry in 3D velocity space offers a principled measure of glide reliability that complements traditional lift-to-drag metrics.

\newpage
\vspace*{-3cm}

\section{Change in Global Phase Space with Orientation: Zimmerman Example} 

\begin{figure}[H]
    \centering
      \vspace*{-0 cm}
    \begin{minipage}[t]{0.56\linewidth}
      \includegraphics[width=\linewidth]{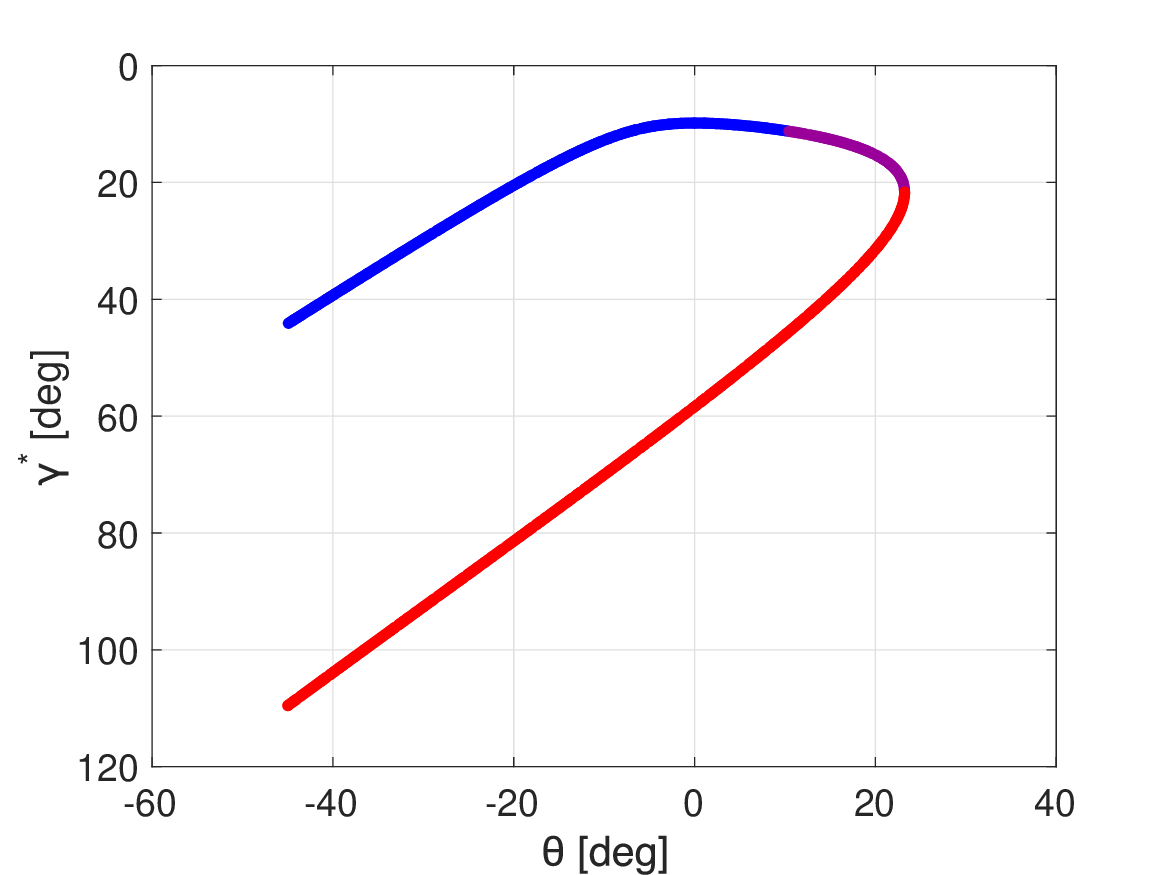}
    \end{minipage}%
    \hfill
    \begin{minipage}[t]{0.56\linewidth}
      \includegraphics[width=\linewidth]{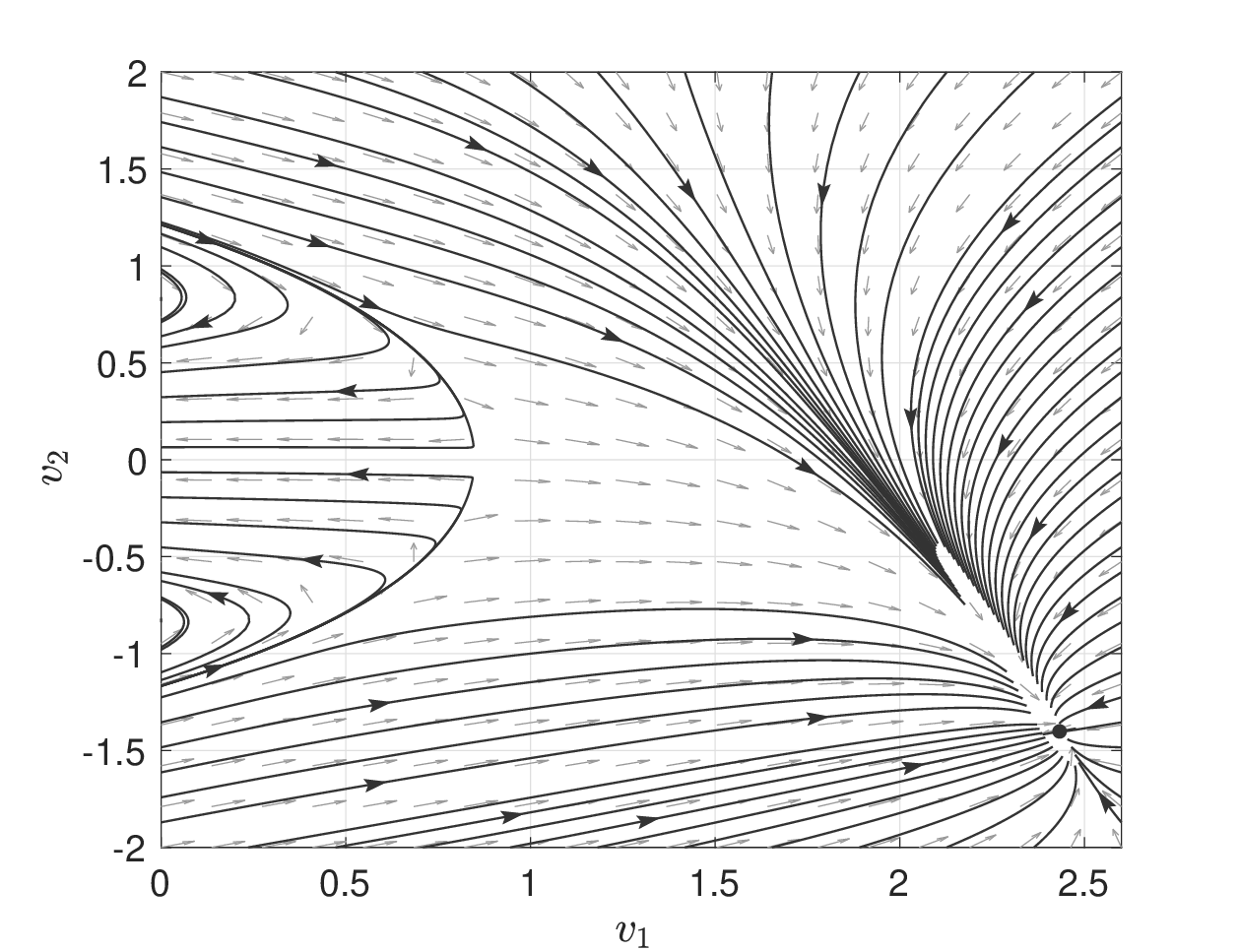}
      \includegraphics[width=\linewidth]{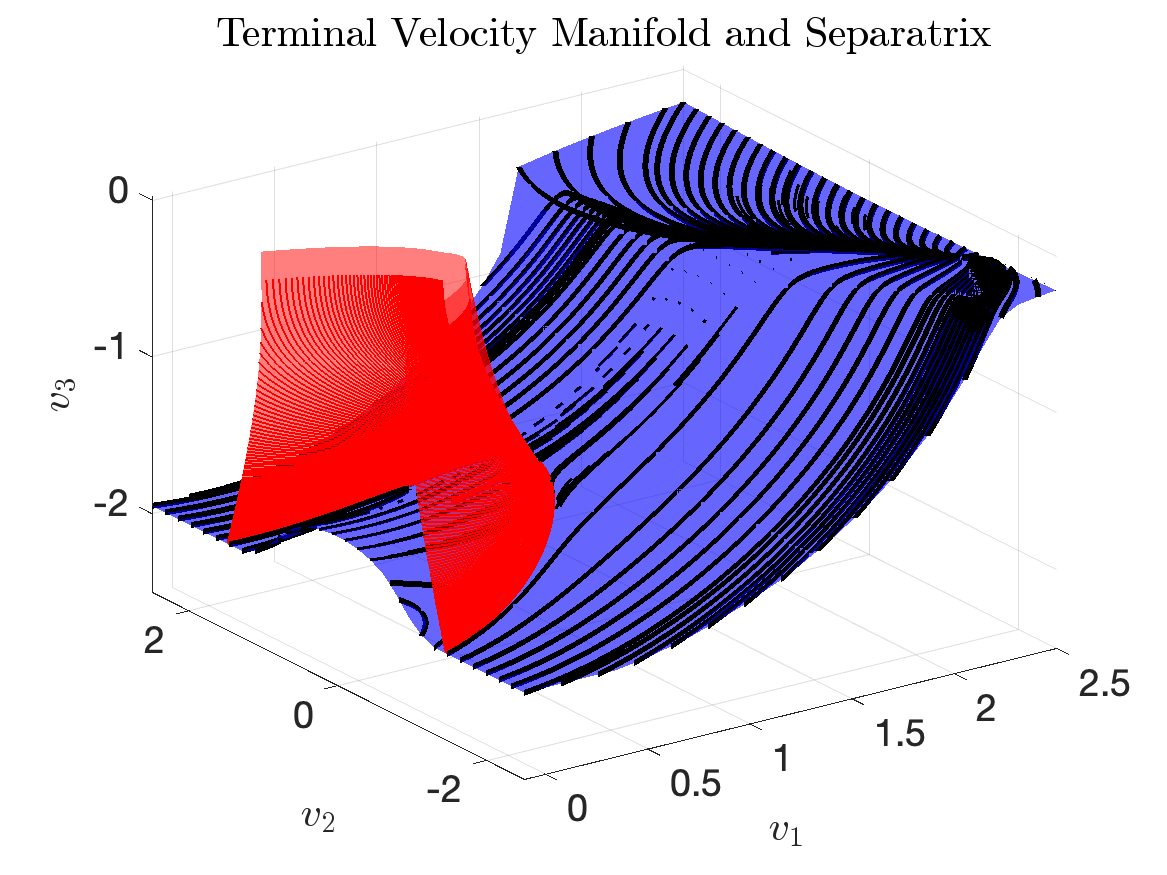} 
    \end{minipage} 
    \label{fig:temporal_NS}
  \caption{(a) Bifurcation diagram of the Zimmerman airfoil at a roll angle of $\phi = 5^\circ$;
(b) the corresponding terminal velocity manifold (TVM);
(c) the separatrix on the TVM distinguishing shallow-glide and steep-descent basins.}
  \label{fig:Zimmerman_roll}
\end{figure}

To illustrate how body orientation reshapes the global invariant structures of the glider, we examine the Zimmerman airfoil under a small roll perturbation.
Figure~\ref{fig:Zimmerman_roll} summarizes how the equilibrium set, the TVM, and the separatrix deform when the roll angle is increased from $\phi = 0^\circ$ (as in Figure~\ref{fig:TVM_combined}(c)) to $\phi = 5^\circ$.
Because the Zimmerman planform is the most robust among the three airfoils studied, it provides a clean example of how orientation shifts the geometry of descent without destroying the underlying invariant structures.

\paragraph{Equilibrium Shift Under Roll.}
The upper panel shows the $(\theta,\gamma^*)$ bifurcation slice at $\phi = 5^\circ$.
Relative to the baseline $(\theta,\phi) = (0^\circ,0^\circ)$ case, the equilibrium branch is displaced both vertically and horizontally: the glide angle $\gamma^*$ increases slightly (a steeper descent) from about 8.5$^\circ$ to 10$^\circ$ (see Figure~\ref{fig:Zimmerman_eq}) and the locus of equilibria shifts toward higher~$\theta$.
Even for this mild roll, the azimuthal angle goes zero to about $-$30$^\circ$ (see Figure~\ref{fig:Zimmerman_eq}).
This reflects how roll introduces a  redistribution of lift and drag along the span, biasing the equilibrium toward higher pitch for the same glide configuration.
Dynamically, trajectories that once converged to the original equilibrium along the $(v_1,v_3)$ plane now drift along the TVM to a new, shifted fixed point corresponding to a gliding descent spiral to the right.

\paragraph{Persistence of the Terminal Velocity Manifold.}
The middle panel shows the TVM at $\phi = 5^\circ$.
The manifold remains smooth and intact, and trajectories still exhibit the hallmark slow-fast structure: rapid collapse onto the manifold followed by slow evolution toward the equilibrium.
The primary change is that the equilibrium point has migrated along the ridge of the TVM, altering the direction of the slow drift.
This introduces a lateral bias in the velocity evolution, demonstrating how roll couples into glide direction even without altering the topology of the manifold.
Trajectories exhibit a lateral bias, an intrinsic coupling between roll and glide direction that alters the path curvature in velocity space.

\paragraph{Deformation of the Separatrix.}
The bottom panel shows the separatrix superimposed on the perturbed TVM.
Relative to the $\phi = 0^\circ$ case, the separatrix tilts and shifts, but not nearly as much as the shallow glide equilibrium. 
This deformation slightly redistributes the basins of attraction: the region of the TVM leading to the shallow, lift-dominated glide becomes slightly smaller, while the drag-dominated basin expands on the opposite side.
In physical terms, roll misalignment increases the tendency toward steeper descent, though the separatrix remains compact enough that shallow-glide trajectories are still readily attainable.

\paragraph{Summary of Orientation Effects.}
Overall, the Zimmerman planform preserves the integrity of its global invariant structures in velocity space under moderate roll perturbation.
The equilibrium shifts smoothly, the TVM retains its topology, and the separatrix deforms but does not fragment or develop new folds.
This structural resilience explains the aerodynamic robustness of the Zimmerman: even at $\phi = 5^\circ$, it maintains predictable, high-efficiency glide dynamics, with smooth transitions between equilibria and no abrupt bifurcations.
The example highlights how orientation reshapes the entire global phase portrait,
even when the underlying dynamical skeleton remains structurally preserved.

\begin{figure}[H]

  \centering
      \includegraphics[width=0.7\linewidth]{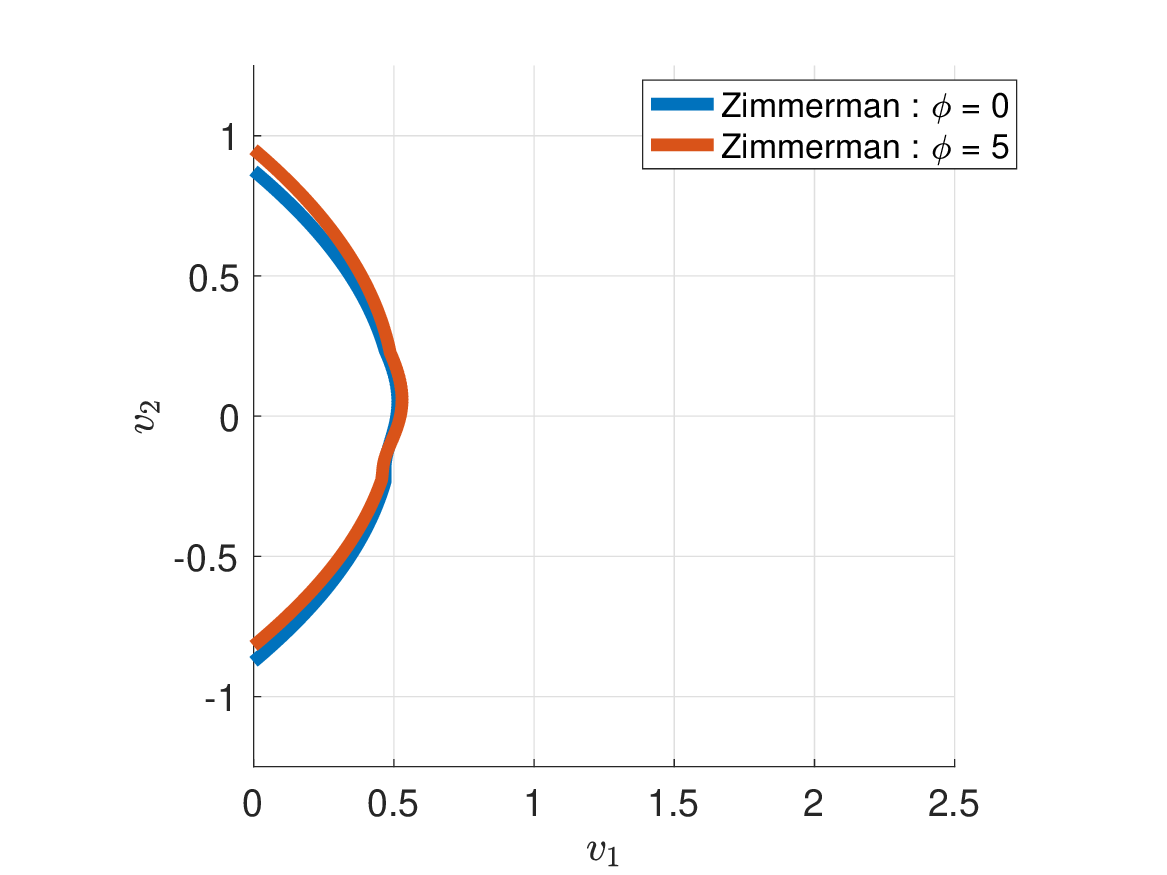} 
  \caption{Separatrix of the Zimmerman planform for two cases: $\phi = 0$ and $\phi = 5$ degrees  }
  \label{fig:sep_zim_case}
  
  \end{figure}
  

\section{Discussion and Conclusions}

The present work extends the classical two-dimensional glider model of Yeaton et al.\ \cite{yeaton2017global} and its subsequent manifold-based refinements into a fully three-dimensional dynamical framework.
By incorporating pitch, roll, and yaw as orientation parameters and by analyzing the resulting phase-space geometry, we reveal a coherent dynamical structure governing passive descent for a wide class of biological and engineered airfoils.

A central finding is the persistence of a two-dimensional terminal velocity manifold (TVM) that attracts all trajectories in  the three-dimensional velocity space.
This surface generalizes the classical notion of terminal velocity from a single vertical speed to an invariant manifold that organizes the velocity space.
Trajectories collapse rapidly onto this two-dimensional slow manifold and then proceed at a much reduced pace along it until reaching an equilibrium glide state.
The TVM therefore provides a natural reduced phase-space skeleton for analyzing glide dynamics, independent of airfoil shape.
Although the TVM depends on airfoil shape and orientation, it persists across all profiles and throughout a broad orientation range, showing that slow-fast organization is a robust feature of non-equilibrium gliding.

The equilibrium states embedded in the TVM depend sensitively on  orientation.
Both pitch and roll can shift equilibrium locations and alter their stability type, producing transitions between node, focus, and saddle behavior.
The snake and Zimmerman airfoils exhibit several equilibrium branches that remain well separated in the full $(\theta,\phi,\gamma^\ast)$ space despite appearing to overlap in projection, while the NACA 0012 displays pronounced multistability with multiple competing equilibria and alternating stability.
These differences arise directly from the nonlinear lift and drag force laws for each profile.


The stable manifold of the steep-descent saddle equilibrium forms a separatrix: a codimension-one invariant surface attached to the TVM that divides shallow, lift-dominated glide trajectories from steep, drag-dominated descent---acting as a binary switch separating gliding from falling.
Its intersection with the horizontal plane $v_3 = 0$ provides a clear diagnostic for {\it which} horizontal jump conditions lead to efficient glide versus those that fall irreversibly toward steeper descent.
The geometry of this separatrix varies markedly across the three airfoils.
The engineered NACA profile exhibits a broad and displaced separatrix, indicating that only large horizontal initial speeds lead to gliding.
In contrast, both bio-inspired profiles---the snake and {\it Draco}/Zimmerman---possess relatively compact separatrix regions of small initial horizontal speeds.
Smaller initial horizontal speeds mean less energy needed to let the natural dynamics lead to the shallow gliding, ensuring that a large set of initial velocities safely converges to efficient descent.
The Zimmerman airfoil is particularly robust: its separatrix is centrally located on the TVM and is well separated from the shallow-glide branch, reflecting its high lift-to-drag ratio and gentle stall behavior.

These results highlight that glide performance cannot be assessed solely by equilibrium glide angles or force-balance considerations.
The geometry of global invariant structure---the TVM, the equilibrium branches, and especially the separatrix---plays a decisive role in determining whether efficient glide states are dynamically attainable from realistic initial conditions.
Biological gliders appear to exploit this principle: their airfoils produce phase-space structures with large basins of attraction surrounding shallow glides, suggesting an evolutionary premium on robustness to perturbations and variability in launch direction.
Engineered profiles such as the NACA 0012 may attain favorable glide ratios under idealized conditions, but their broader separatrix regions indicate that substantially more energy---or more precisely, a more carefully aligned initial velocity---is required to enter the efficient glide state.
This is consistent with the design intent of the NACA 0012: as Abbott \& von Doenhoff \cite{abbott1949theory} describe, NACA sections were developed for aircraft wings operating at high Reynolds numbers, small angles of attack, and high lift-to-drag ratios in cruise, rather than for the low-speed equilibrium-descent regimes characteristic of passive gliding \cite{khandelwal2023convergence}.

The present analysis has been carried out under several simplifying assumptions, including rigid-body shape, prescribed orientation, and quasi-steady lift and drag.
These assumptions allow a clear characterization of the phase-space geometry but do not capture unsteady aerodynamics, passive deformation, or attitude dynamics.
Incorporating these effects into a fully coupled model would produce richer bifurcation and control structure, but the global geometric framework developed here, especially the role of the TVM and separatrix surfaces, should remain foundational.

In conclusion, this three-dimensional extension of the glider model reveals a unified dynamical-systems organization underlying passive descent across disparate airfoils.
The comparative analysis of snake, NACA, and Zimmerman profiles shows how nonlinear aerodynamic force laws shape the global structure of equilibria and separatrices, ultimately determining glide robustness.
These insights provide a dynamical foundation for understanding biological gliding strategies and suggest principled design criteria for future bio-inspired aerial vehicles.


\section{Acknowledgements}

This work was supported by the National Science Foundation under Grant No. 2027523 to S.D.R.

\section{Author contributions}
M.Z. and S.D. designed the study, interpreted the results, and contributed to the writing of the manuscript.
M.Z. and S.D.  implemented the models, conducted simulations, analyzed the data, and produced the figures. All authors reviewed the manuscript.

\section*{Appendix. Derivations}

Below is the most general expression for the inertial-frame unit velocity vector $\hat{\mathbf{v}}$ in terms of the five `body-relative' angles $\phi$ (roll), $\theta$ (pitch), $\psi$ (yaw), $\alpha$ (angle of attack) and $\beta$ (sideslip), 
followed by the three equations one obtains by equating those components to the `flight-path' description.
In the NED inertial frame convention (not shown in first figure),
\begin{equation} 
^{\mathcal{N}}\hat{\mathbf{v}} \;=\; \bigl(\cos\gamma\,\cos\sigma,\;-\cos\gamma\,\sin\sigma,\;\sin\gamma\bigr)\,.
\end{equation}

1. Full, general expression for $^{\mathcal{N}}\hat{\mathbf{v}}$
Starting from
\begin{equation}   
[BN] =
\begin{bmatrix}
\cos\psi \cos\theta & \sin\psi \cos\theta & -\sin\theta \\
\cos\psi \sin\theta \sin\phi - \cos\phi \sin\psi &
\sin\psi \sin\theta \sin\phi + \cos\phi \cos\psi &
\cos\theta \sin\phi \\
\cos\phi \sin\theta \cos\psi + \sin\phi \sin\psi &
\sin\psi \sin\theta \cos\phi - \sin\phi \cos\psi &
\cos\theta \cos\phi
\end{bmatrix} 
\end{equation}
and
\begin{equation}  
[WB] =
\begin{bmatrix}
\phantom{-}\cos\alpha\,\cos\beta & \;\sin\beta & \;\sin\alpha\,\cos\beta \\
-\cos\alpha\,\sin\beta & \;\cos\beta & -\,\sin\alpha\,\sin\beta \\
-\,\sin\alpha & \;0 & \;\cos\alpha
\end{bmatrix}
\end{equation}   
then the wind frame  to inertial frame rotation is $[NW] = [BN]^T [WB]^T$, and $^{\mathcal{N}}\hat{\mathbf{v}}$ can be related to $^{\mathcal{W}}\hat{\mathbf{v}} = (1, 0, 0)^T$ via
\begin{equation}    
^{\mathcal{N}}\hat{\mathbf{v}}
= [NW] ^{\mathcal{W}}\hat{\mathbf{v}}.
\end{equation}
When you carry out those two transposes and multiply by $\bigl(1,0,0\bigr)^{T}$, you get, component-wise,
\begin{equation}   
\tiny
^{\mathcal{N}}\hat{\mathbf{v}}
=
\begin{bmatrix}
\bigl(\sin\phi \sin\psi + \sin\theta \cos\phi \cos\psi \bigr) \sin\alpha \cos\beta
+\bigl(\sin\phi \sin\theta \cos\psi  - \sin\psi \cos\phi \bigr) \sin\beta
+\cos\alpha \cos\beta \cos\psi \cos \theta
\\[1.25em]
-\bigl(\sin\phi \cos\psi  - \sin\psi \sin\theta \cos\phi\bigr) \sin\alpha \cos\beta
 + \bigl(\sin\phi \sin\psi \sin\theta +\cos\phi \cos\psi \bigr) \sin\beta
 + \cos\alpha \cos\beta \sin\psi \cos\theta
\\[1.25em]
\cos\phi \cos\theta \sin\alpha \cos\beta
+\sin\phi \cos\theta \sin\beta
-\sin\theta \cos\alpha \cos\beta
\end{bmatrix}.
\end{equation}

2. Equate to the flight-path form

We also know, by definition of $\gamma$ (vertical flight-path angle) and $\sigma$ (horizontal azimuth), that in the same inertial frame
$^{\mathcal{N}}\hat{\mathbf{v}} = (
\cos\gamma \cos\sigma, 
-\cos\gamma \sin\sigma,
\sin\gamma)^T$
Hence each component must match.

In other words, the three scalar equations are
\begin{equation}
\tiny
    \begin{aligned}
(1)\quad
&\bigl(\sin\phi\,\sin\psi + \sin\theta\,\cos\phi\,\cos\psi\bigr)\,\sin\alpha\,\cos\beta
\;+\;\bigl(\sin\phi\,\sin\theta\,\cos\psi - \sin\psi\,\cos\phi\bigr)\,\sin\beta
\;+\;\cos\alpha\,\cos\beta\,\cos\psi\,\cos\theta
\;=\;\cos\gamma\,\cos\sigma,
\\[0.75em]
(2)\quad
-&\bigl(\sin\phi\,\cos\psi - \sin\psi\,\sin\theta\,\cos\phi\bigr)\,\sin\alpha\,\cos\beta
\;+\;\bigl(\sin\phi\,\sin\psi\,\sin\theta + \cos\phi\,\cos\psi\bigr)\,\sin\beta
\;+\;\cos\alpha\,\cos\beta\,\sin\psi\,\cos\theta
\;=\;-\,\cos\gamma\,\sin\sigma,
\\[0.75em]
(3)\quad
&\cos\phi\,\cos\theta\,\sin\alpha\,\cos\beta
\;+\;\sin\phi\,\cos\theta\,\sin\beta
\;-\;\sin\theta\,\cos\alpha\,\cos\beta
\;=\;\sin\gamma.
\end{aligned}
\end{equation}
These three equations are the fully general form of
$\hat{\mathbf{v}}^{\,[B]} \;\longrightarrow\; \hat{\mathbf{v}}^{\,[N]}$,
with zero assumptions removed.  Solving this system simultaneously gives $\alpha$, $\beta$, and $\theta$ (and implicitly $\phi$, $\psi$) in terms of $\gamma$, $\sigma$ and the body-orientation angles.

Notes
	1.	If you later impose $\phi=0$ (level wings) or $\psi=0$ (no yaw) or any other special cases, you simply set those symbols to zero in the three equations above.
	2.	Each line is one inertial component equation (x, y, or z).  In practice, one often solves (3) for $\alpha$ (given $\beta,\phi,\theta$), and then uses (1) and (2) to pin down $\beta$ (given $\gamma,\sigma,\phi,\psi,\theta$).
	3.	In many textbooks you will see the level-flight simplification $\phi=0$, $\psi=0$, which immediately collapses these to   
$\begin{cases}
\cos\theta\,\sin\alpha\,\cos\beta - \sin\theta\,\cos\alpha\,\cos\beta \;=\;\sin\gamma,\\
\cos\alpha\,\cos\beta\,\cos\theta = \cos\gamma\,\cos\sigma,\\
\quad-\cos\alpha\,\cos\beta\,\sin\theta = -\,\cos\gamma\,\sin\sigma.
\end{cases}$

\section*{Appendix. Jacobian}

The equation of motion in compact form is,
\begin{equation}
\dot{\vv} = \mathbf{A}(\vv) \vv - \hat{\nn}_3,
\end{equation}
The aerodynamic component of the dynamics is represented by, $\mathbf{A}(\vv)$, can be decomposed into a drag and lift component,
\begin{equation}
\mathbf{A}(\vv)
= \mathbf{D}(\vv) + \mathbf{L}(\vv)
=
-\,v^{}\,C_{D}(\alpha^{})\,\mathbf{I}_3
\;+\;v^{}\,C_{L}(\alpha^{})\,
\tilde{\mathbf{L}},
\end{equation}
where
\begin{equation}
\tilde{\mathbf{L}} = 
\begin{bmatrix}
0 & \phi & -\,a_{\phi\psi}\\
-\,\phi & 0 & a_{\psi\theta\phi}\\
a_{\phi\psi} & -\,a_{\psi\theta\phi} & 0
\end{bmatrix}.
\end{equation}
We note that  $\tilde{\mathbf{L}} \vv$ can be written as a cross-product,
\begin{equation}
\tilde{\mathbf{L}} \vv = -\ttt_{L} \times \vv,
\end{equation}
where the vector $\ttt_L$ depends on Euler angles,
\begin{equation}
    \ttt_{L} =
    \begin{bmatrix}
    a_{\psi\theta\phi} \\
    a_{\phi\psi} \\
    \phi
    \end{bmatrix}.
\end{equation}
So the equation of motion can be written as,
\begin{equation}
\dot{\vv} = -v C_D(\vv) \vv - v C_L(\vv)\ttt_{L} \times \vv - \hat{\nn}_3,
\end{equation}

The gradient $\nabla(v\,C_{D})\bigl|_{\vv^{*}}$ is,
\begin{equation}
\nabla\bigl(v\,C_D(\alpha(\vv))\bigr)
\;=\;
\begin{bmatrix}
\frac{\partial}{\partial v_{1}}\bigl(v\,C_{D}\bigr) \\[6pt]
\frac{\partial}{\partial v_{2}}\bigl(v\,C_{D}\bigr) \\[6pt]
\frac{\partial}{\partial v_{3}}\bigl(v\,C_{D}\bigr)
\end{bmatrix},
\end{equation}
evaluated at $\vv^*$.  The dot product $\nabla(v\,C_{D})\bigl|_{\vv^{*}}\cdot \vv^{*}$ is a scalar, and it multiplies the 3$\times$3 identity $\mathbf{I}_{3}$.

The gradient $\nabla(v\,C_{L})\bigl|_{\vv^*}$
is defined analogously.  Its dot product with $\vv$ is another scalar, which then multiplies the fixed skew-symmetric matrix $\tilde{\mathbf{L}}$.
Because each of the three terms, $A(\vv^*)$, $-\bigl[\nabla(v\,C_{D})\cdot \vv^*\bigr]\,\mathbf{I}_3$, and $\bigl[\nabla(v\,C_{L})\cdot \vv^*\bigr]\,\tilde{\mathbf{L}}$, is now explicit this completes a self-contained expression for the Jacobian.

\section{Declarations}
\subsection*{Conflict of Interest}
The authors declare that they have no known competing financial interests or personal relationships that could have appeared to influence the work reported in this paper.

\section{Availability of Data and Materials}
The code used to generate the results in this study is available from the corresponding author upon reasonable request.

\end{document}